\documentclass[aps,twocolumn,showpacs,prc,floatfix]{revtex4}
\usepackage{graphicx}
\usepackage{epsf}
\usepackage{wrapfig}
\usepackage{epsfig}

\def\b0{{\mbox{\boldmath$0$}}}

\usepackage{graphicx}
\usepackage{epsf}
\usepackage{wrapfig}
\usepackage{epsfig}
\usepackage{sublabel}
\usepackage{epsfig}
\usepackage{epstopdf}
\def\Vec#1{\mbox{\boldmath $#1$}}

\def\beq{\begin{equation}}
\def\eeq{\end{equation}}
\def\nn{\nonumber}
\def\beqy{\begin{eqnarray}}
\def\eeqy{\end{eqnarray}}

\def \b #1{ {\bf #1}}
\newcommand{\be}{\begin{eqnarray}}
\newcommand{\ee}{\end{eqnarray}}

\def \b #1{ {\bf #1}}

\def \b #1{ {\bf #1}}
     \font\tenbifull=cmmib10 scaled 1200 
     \font\tenbimed=cmmib9
     \font\tenbismall=cmmib7
\mathchardef\bbkappa="7114
\mathchardef\bbrho="711A
\mathchardef\bbsigma="711B
\mathchardef\bbtau="711C
\mathchardef\bbvarrho="7125
\mathchardef\bbvarsigma="7126
\mathchardef\bbxi="7118

\begin{document}
\vskip 2mm \date{\today}\vskip 2mm
\title{Beam Fragmentation in Heavy Ion Collisions with Realistically Correlated Nuclear Configurations}
\author{M. Alvioli\footnote{
    Present address: ECT$^*$, Strada delle Tabarelle 286, I-38123, Villazzano (TN) Italy}, M. Strikman}
\affiliation{The Pennsylvania State University, 104 Davey Lab, 
  University Park, Pennsylvania 16803, USA}
 \vskip 2mm
\begin{abstract}
We develop a new approach to production of the spectator nucleons in the ultra-relativistic
heavy ion collisions.
The energy transfer to the spectator system is calculated using the Monte Carlo based on the
updated version of our generator of configurations in colliding nuclei which includes a realistic
account of short-range correlations in nuclei. The transferred energy distributions are calculated
within  the framework of the Glauber multiple scattering theory, taking into account all individual
inelastic and elastic collisions using an independent realistic calculation of the potential energy
contribution of each of the nucleon-nucleon pairs to the total potential. We show that the dominant
mechanism of the energy transfer is tearing apart pairs of nucleons with the major contribution coming
from the short-range correlations. We calculate the momentum distribution of the directed flow of emitted
nucleons which is strongly affected by short range correlations including its dependence on the azimuthal
angle. In particular, we predict a strong angular asymmetry along the direction of the  impact parameter
$\Vec{b}$, providing a unique opportunity to determine the direction of $\Vec{b}$. Also, we  predict a
strong dependence of the shape of the nucleon momentum distribution on the centrality of the nucleus-nucleus
collision.
\end{abstract}
\maketitle

\section{Introduction}\label{sec:intro}

In this paper we start a program of studies of the nuclear fragmentation in ultra-relativistic
heavy ion collisions using as a starting point an event generator
of the nucleon configurations in nuclei which correctly reproduces short-range
correlations between the nucleons.

Most of the recent experimental and theoretical studies of the relativistic heavy
ion collisions were focused on the production of hadrons at central rapidities.
Fragmentation of nuclei in these collisions was used only as a supplementary trigger
for centrality. At the same time experiments at Relativistic Heavy Ion Collider (RHIC) 
have demonstrated that it is
possible to determine on the event by event basis impact parameter and reaction
plane of the collision. (There is obviously some inherit uncertainty related to
the fluctuations of the observables for collisions at a given impact parameter.)
This opens new opportunities for studies of the nuclear fragmentation which have
a long history; see, for example, Ref. \cite{Scheidenberger:2004xq} and references
therein.

Another motivation is the recent direct  observation of the short-range correlations
(SRC)  \cite{Subedi:2008zz,Shneor:2007tu,Piasetzky:2006ai,Tang:2002ww} in the nuclear
decays initiated by a hard removal of the nucleon from the nucleus. When combined with
the scaling of the ratios in $x > 1$ $(e,e^\prime)$ nuclear reactions, it demonstrates
the important role played by SRC in nuclear structure; for a recent review, see
Ref. \cite{Frankfurt:2008zv}. This calls for a description of heavy ion collisions using
realistic configurations of nucleons in nuclei going  beyond the commonly used collection
of nucleons randomly distributed in the nuclear volume.

We started  the program of including SRC in nuclear configurations in Ref. \cite{Alvioli:2009ab}
where we implemented central correlation functions using the Metropolis method. This allowed us
to overcome the problem of distortion of single particle density when configurations with nucleons
at short distance are simply discarded. The pair distribution function for the generated configurations
is close enough to the realistic one; in particular, it vanishes when the two nucleons separation
approaches zero.

In Ref. \cite{Alvioli:2009ab} the inclusion of correlations in nuclear configurations was shown to have
significant effects on the  fluctuations of the average number collisions in $NA$ scattering; the authors
of Ref. \cite{Broniowski:2009fm,Broniowski:2010jd} confirmed the importance of $NN$ correlations
effects on fluctuations in $NA$ as well as $AA$ collisions using our central-correlated configurations
\cite{download} within their Monte Carlo (MC) simulations. In this work we implement the second
step of our program: taking into account spin and isospin dependence in the generation of configurations.
This allows us to implement for the first time, the state-dependent correlations; this procedure is 
discussed in Sec. \ref{sec:mc}.

In this paper we apply the MC method for taking into correlation effects in AA collisions
at the energies of the CERN NA49 experiment \cite{Appelshauser:1998tt},
also using results from the Lawrence Berkeley National Laboratory (LBL) experiment of Ref. 
\cite{Anderson:1982jx}. 
We analyze dependence of nuclear fragmentation on centrality of the collisions.
In the data  
analyzes the centrality is usually determined based on the
the correlation between the observed charged particle multiplicity and the calculated
number of participant nucleons in the $AA$ collision. This
leads to large uncertainties 
and the impact parameter is usually known in large bins and very central collisions are
difficult to identify. We propose in this work 
\begin{figure*}[!htp]
\leftline{\hspace{1cm}
  \epsfxsize=7.5cm\epsfbox{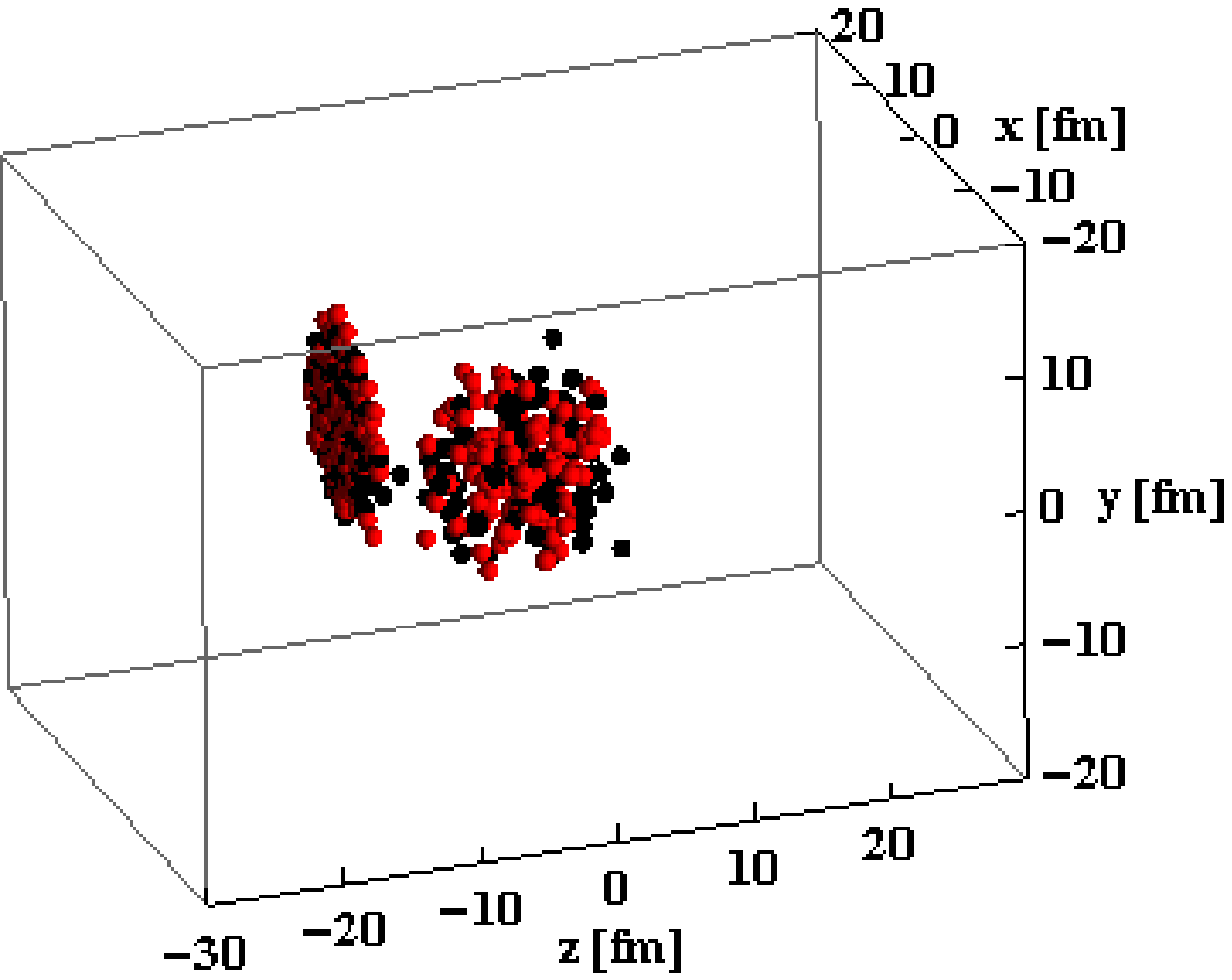}}
\vskip -6.2cm
\leftline{\hspace{9cm}
  \epsfxsize=7.0cm\epsfbox{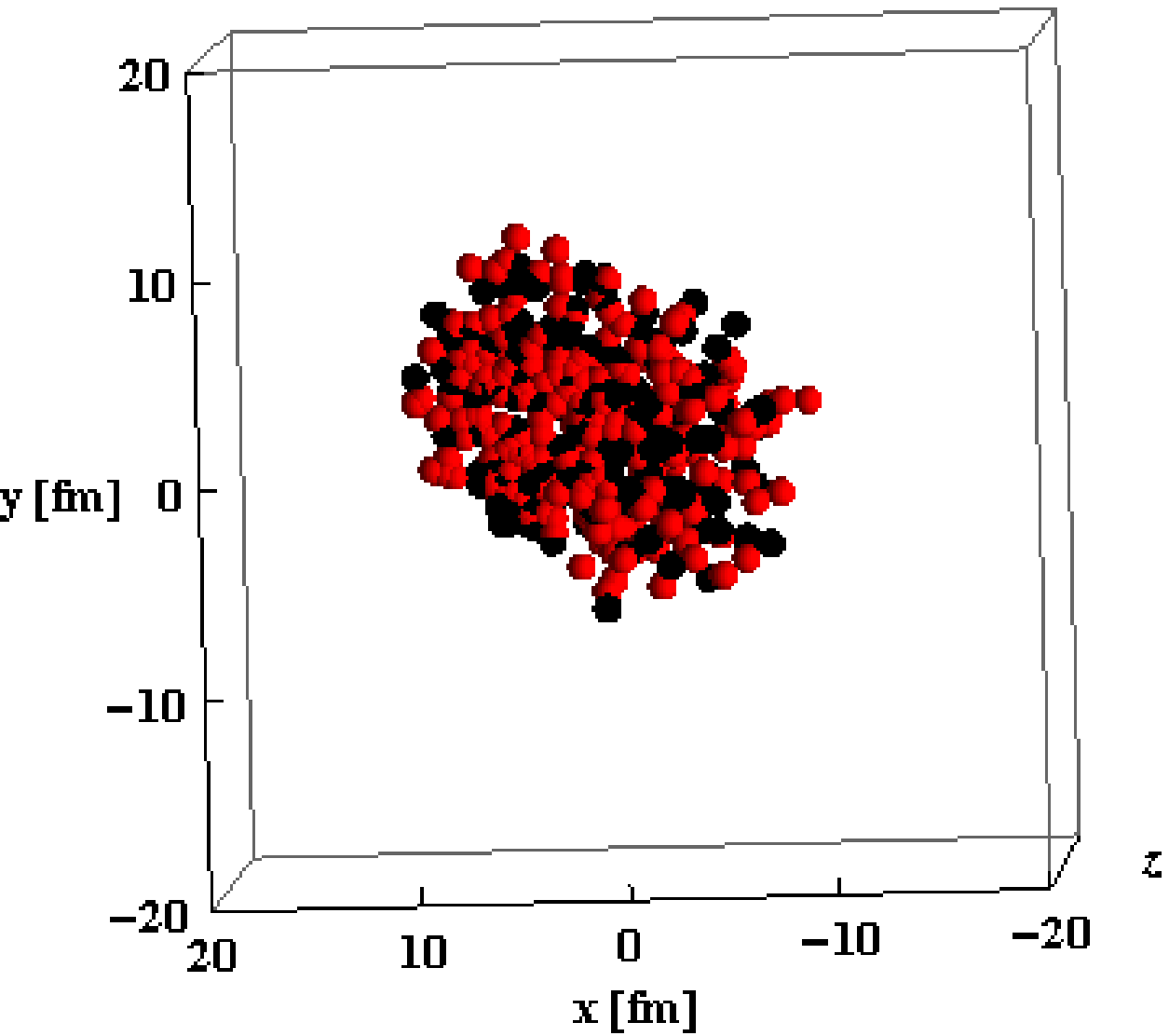}}
\vskip 0.2cm
\leftline{\hspace{1cm}
  \epsfxsize=7.5cm\epsfbox{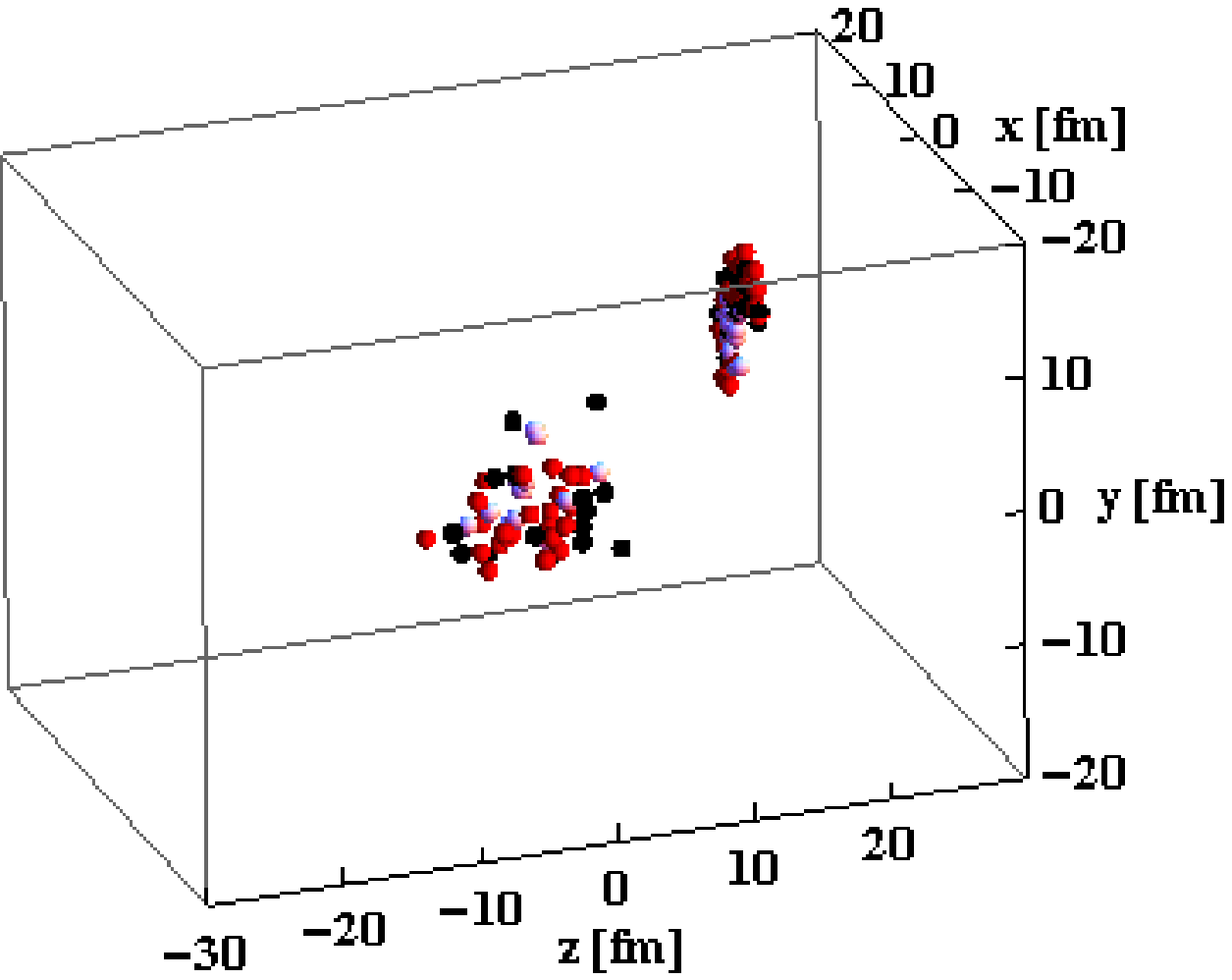}}
\vskip -6.2cm
\leftline{\hspace{9cm}
  \epsfxsize=7.0cm\epsfbox{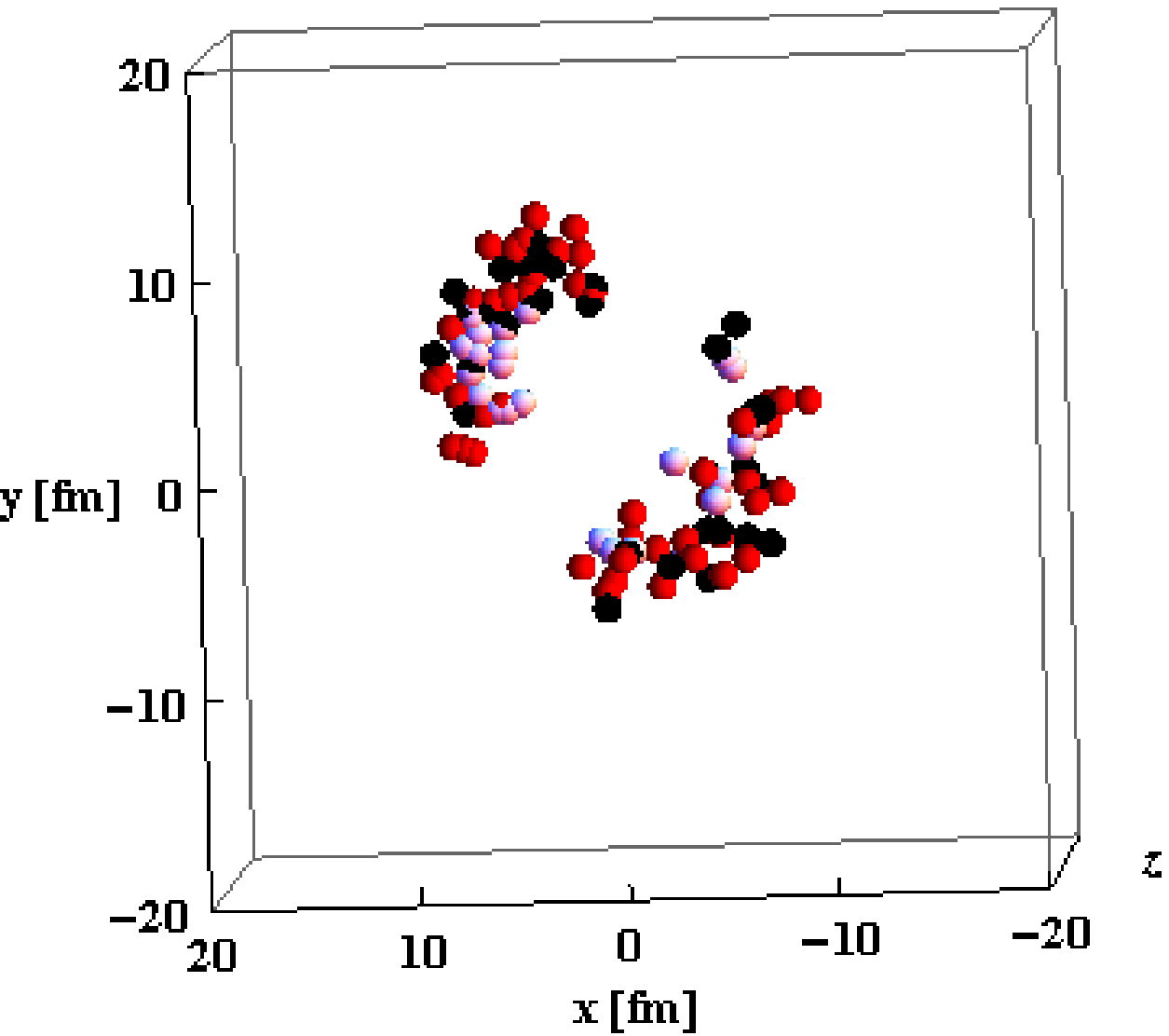}}
\vskip -12.0cm                                   %
\centerline{\large\textbf{(a)}\hspace{12.5cm}\textbf{(b)}\hspace{1.5cm}}
\vskip 5.8cm                                     %
\centerline{\large\textbf{(c)}\hspace{12.5cm}\textbf{(d)}\hspace{1.5cm}}
\vskip 5.3cm                                     %
\caption{(Color online) Sketch of a $Pb-Pb$ collision in the target rest frame at impact
  parameter $b=5$ fm oriented at 45$^o$ with respect to the $x$ and $y$ axes; the
  projectile moves along the $z$ axis. [(a) and (c)] View along the beam line;
  [(b) and (d)] view from behind; [(a) and (b)] before interaction;
  [(c) and (d)] after interaction. The inelastically interacting nucleons
  have been removed from the figure.
  Black and red spheres represent protons and neutron, respectively, while the
  white ones are active nucleons (see text).
  The dimension of the spheres are taken as the rms charge radius of the proton.
  Animations are available at the URL in Ref. \cite{download} along with
  the configurations used for the colliding nuclei.
  }\label{Fig1}
\end{figure*}
that one can obtain additional information
about the centrality of a collision, based on the detailed balance of energy transferred in
the collision at a given impact parameter and on the emission of high-momentum nucleons
originating from SRC pair in nuclei. In addition, the  angular asymmetry for emission of
nucleons allows us to resolve the sign ambiguity of the direction of impact parameter
$\Vec{b}$ in contrast with the current procedures which determine only $\left|\Vec{b}\right|$.

We introduce a new model for the description of the emission mechanism of spectator nucleons
in ultra-high-energy heavy ion collisions.
In this limit, in the rest frame of one of the nuclei, the projectile is strongly Lorentz contracted
to the longitudinal size of the order $\le 1/\mu$ where $\mu $ is a soft strong interaction scale
$\le m_{\rho}$. As a result, the collision can be considered as the propagation of a ``pancake''
moving with the speed of light which consequently removes  nucleons from the target along its path;
the situation is depicted in Fig.~\ref{Fig1}.
It is worth emphasizing that the dynamics of nuclear fragmentation at low energies differens
significantly; energy transfers to individual nucleons are small as compared to the scale
of energies in SRC, and the relative velocities of nuclei are small. This differs markedly
from the picture of a thin pancaked nucleus going through the target nucleus we can employ
in the ultra-relativistic collisions. For a review of models of fragmentation at low energies,
see Ref. \cite{Aichelin:1991xy}.
The spectator system emerging from the target is left in an unknown excited state. At the first stage
the destruction of the potentials between the pairs (triplets) of nearby nucleons, one of which is hit
and another belongs to the spectator system, release energy on the layer of nucleons adjacent to the
removed portion of the nucleus. This energy comes from the work performed by destruction of the potential
energy (bonds) associated with the position of these nucleons in the initial wave function, and it can be
readily converted into kinetic energy. As a result, they can be emitted into free space in the direction
\begin{figure*}[!htp]
\vskip -0.2cm
\centerline{\includegraphics[height=6.5cm]{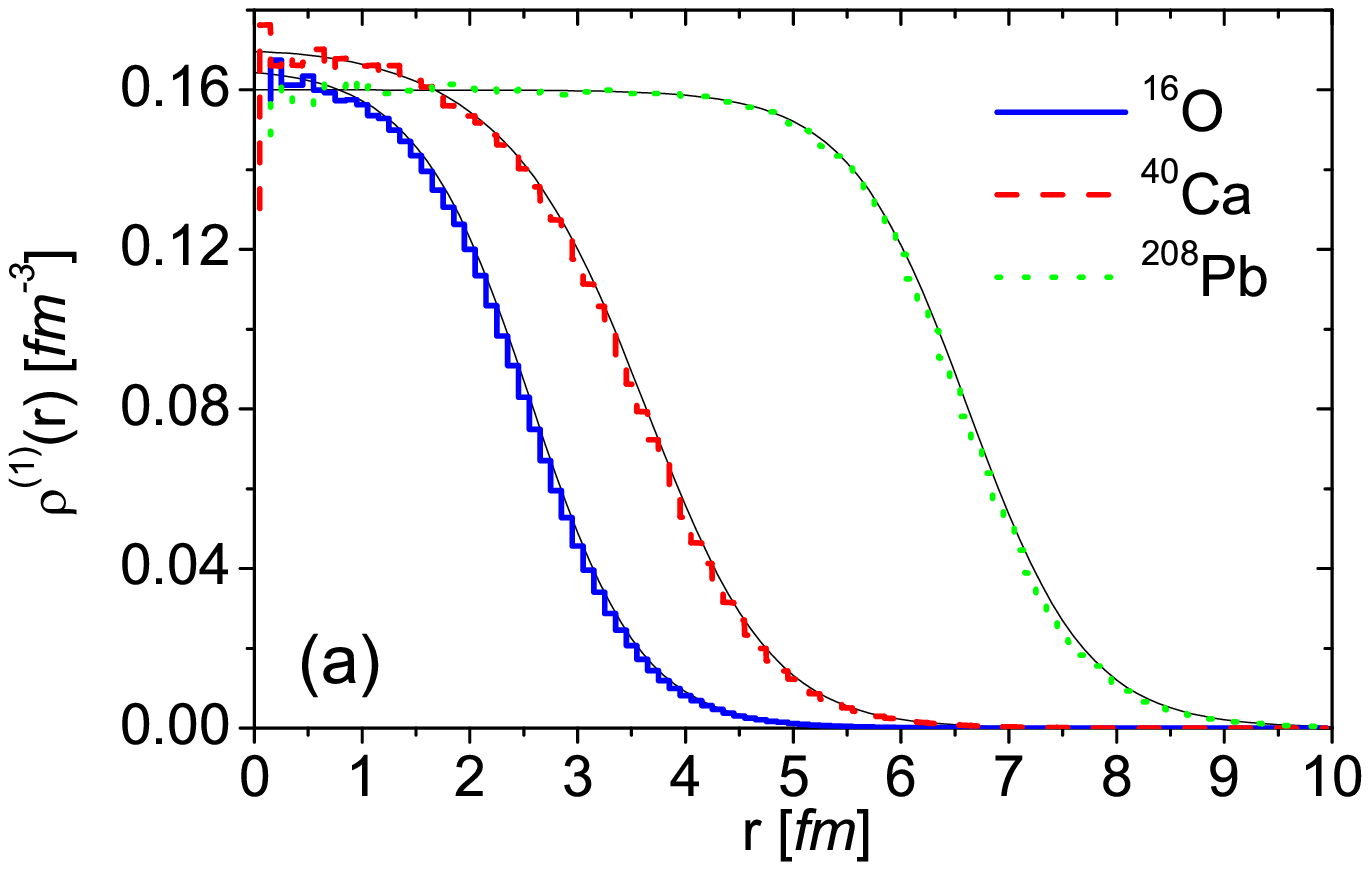}
\hspace{-0.5cm}\includegraphics[height=6.5cm]{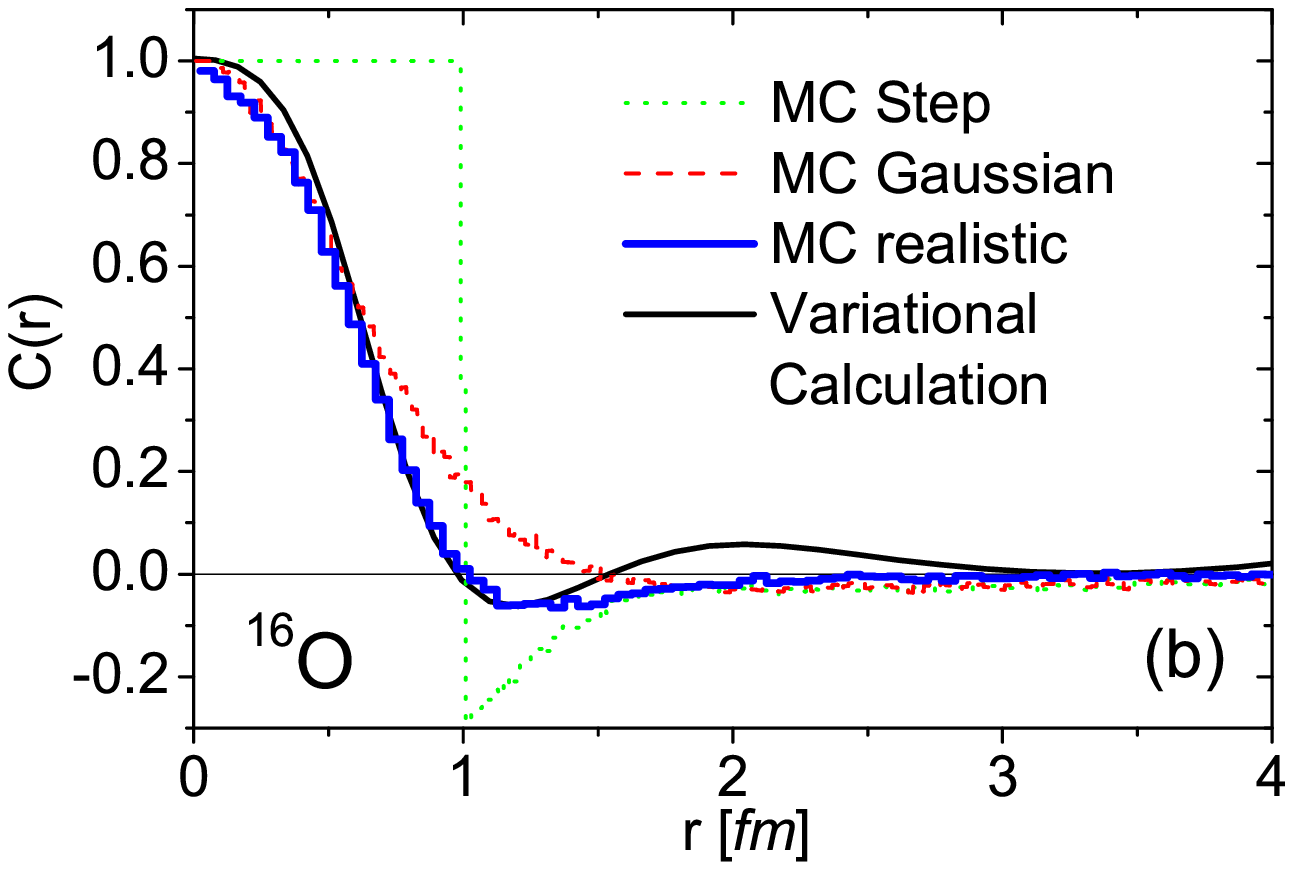}}
\vskip -0.2cm
\caption{(Color online) (a) The one-body density of $^{16}O$, $^{40}Ca$
  and $^{208}Pb$; the colored curves correspond to the calculation with our fully correlated
  configurations, while the black curves are the corresponding input functions.
  (b) The pair distribution function $C(r)$ of $^{16}O$ defined in Eq.
  (\ref{eqpair}), obtained with MC within various approximations for the nucleon-nucleon
  correlation functions contained in Eq. (\ref{defpsi}), compared with the variational calculation
  of Ref. \cite{Alvioli:2005cz}. Calculations for $^{40}Ca$ and $^{208}Pb$ nuclei produce
  similar results.}
\label{Fig2}
\end{figure*}
of the removed nucleons, or propagate into the spectator system and undergo attenuation. The remaining
available potential energy goes into excitation of the spectator system and it can be released at a later
stage by standard nucleon evaporation and decay. Experimental analyses along the lines we suggest in this
work are feasible at RHIC as presence of the directed flow $v_1$ for neutrons was observed (though not
analyzed in detail) by the PHENIX \cite{Afanasiev:2009wq} and STAR \cite{Adams:2005ca,Abelev:2008jga}
experiments.

The paper is organized as follows.
In Sec. \ref{sec:mc} we introduce general definitions and describe our improved procedure for generating 
nucleon configurations in nuclei, which, in difference from our original procedure, includes spin and 
isospin nucleon-nucleon correlations.  We also describe the MC procedure for  the first stage of the 
collision which separates nucleons into spectators and interacting nucleons. It employs the geometry 
of the Glauber description of the collision, which allows an impact-parameter-dependent description 
and takes into account the basic features of realistic calculations of SRCs in the ground state
of nuclei \cite{Alvioli:2005cz}. (Note here that the Glauber model forms a basis of many MC codes
for the simulation of high energy nucleus-nucleus collisions; for example, HIJING \cite{Wang:1991hta}
for the simulation of nucleus-nucleus collisions.)  In our model we use the standard description of
hadron-nucleus interactions \cite{Bialas:1976ed}, in which inelastic collisions of nuclei are treated
as an incoherent superposition of the
individual collisions of the nucleons of the two systems. The  participants are defined as the
nucleons which interacted inelastically at least once in the $AA$ collision, and spectators are the
nucleons which did not interact. They are determined within the Glauber approximation and the
simulation of the collisions  is performed   starting from random nucleons distributed according
to a given probability density for the nucleus profile.

In Sec. \ref{sec:pot} we perform first dynamical calculation of the total excitation energy of the
spectator system based on the total potential energy due to the links between spectators and interacting
nucleons and use of the analog of the Koltun sum rule. Inclusion of SRC is critical at this point.
Indeed, the realistic calculations of the fractions of total potential energy due to the different
$pp$ and $pn$ pairs in nuclei based on the method of Ref. \cite{Alvioli:2005cz} show that, while
in the mean field approximation the two contributions are proportional to the corresponding number
of pairs, the inclusion of correlation drastically changes these fractions, bringing the $pn$ pairs
to carry about $83\%$ of the total potential energy. The results are given as a function of the impact
parameter of the $AA$ collision.

In Sec. \ref{sec:momdis} we first  describe the algorithm for the selection of correlated nucleons
among the nucleons close to the collision surface. 
Next, we outline the procedure used to generate emission of both high momentum and low-momentum 
nucleons from the surface. The soft part of the distribution is due to heating and
evaporation of the residual system which includes taking into account extra heating from absorption
of the part of the nucleons emitted from the surface by the bulk of the spectator system.
We describe also an additional contribution to the  nucleon production
due to elastic scattering of nucleons near the surface of the collision which is enhanced for the
emission angle close to $70 \div 90^o$ in the nucleus rest frame. The section concludes with numerical
results for the nucleon momentum distributions. The relative roles of different mechanisms discussed in
the paper is presented as well as  and their dependence on the impact parameter.

In Sec. \ref{sec:angular} we discuss a novel feature of the proposed mechanism of nucleus fragmentation:
the strong angular dependence of the nucleon emission. 
\begin{figure*}[!htp]
\vskip -0.2cm
\centerline{\includegraphics[height=6.5cm]{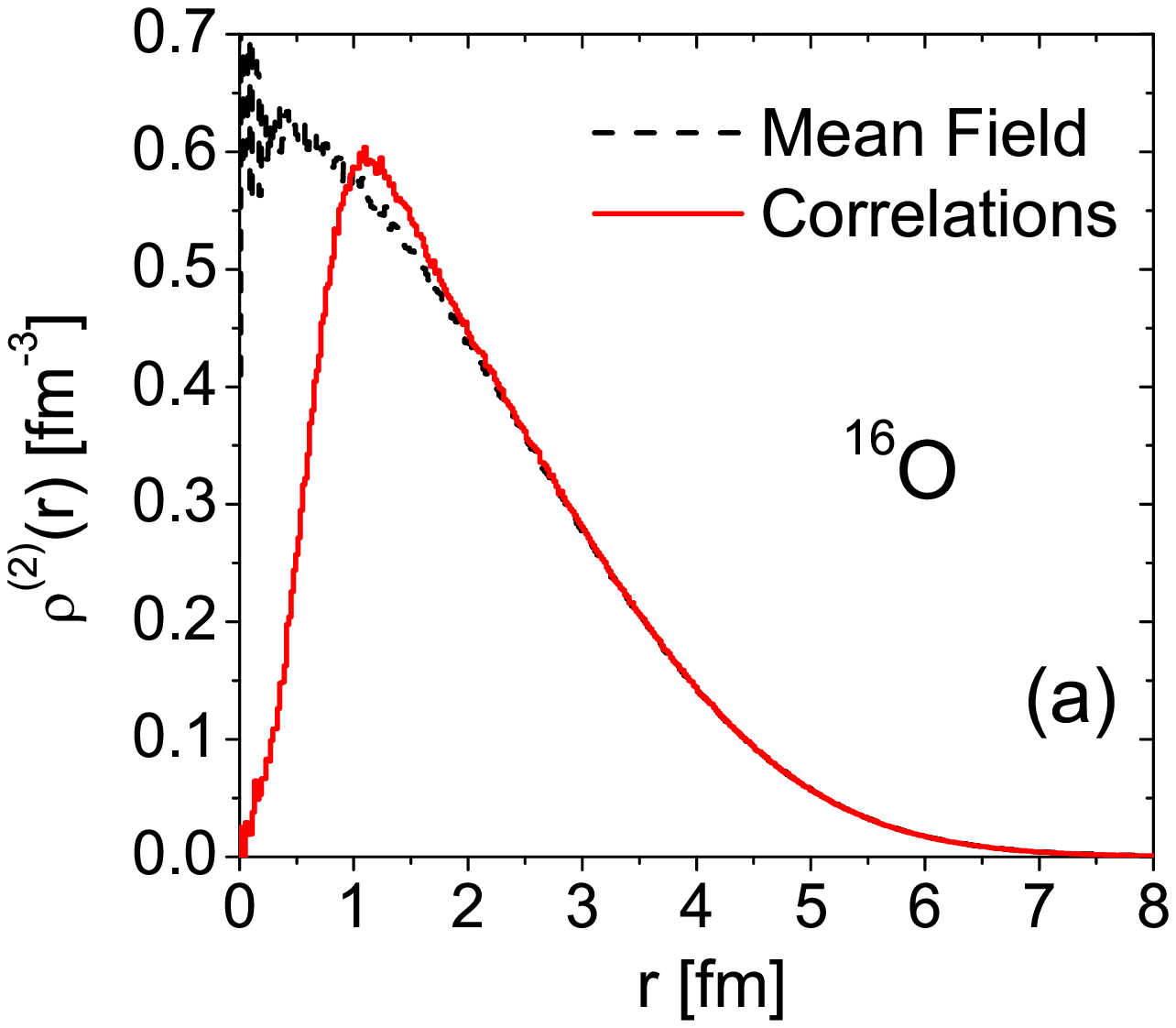}
\hspace{-1.0cm}\includegraphics[height=6.5cm]{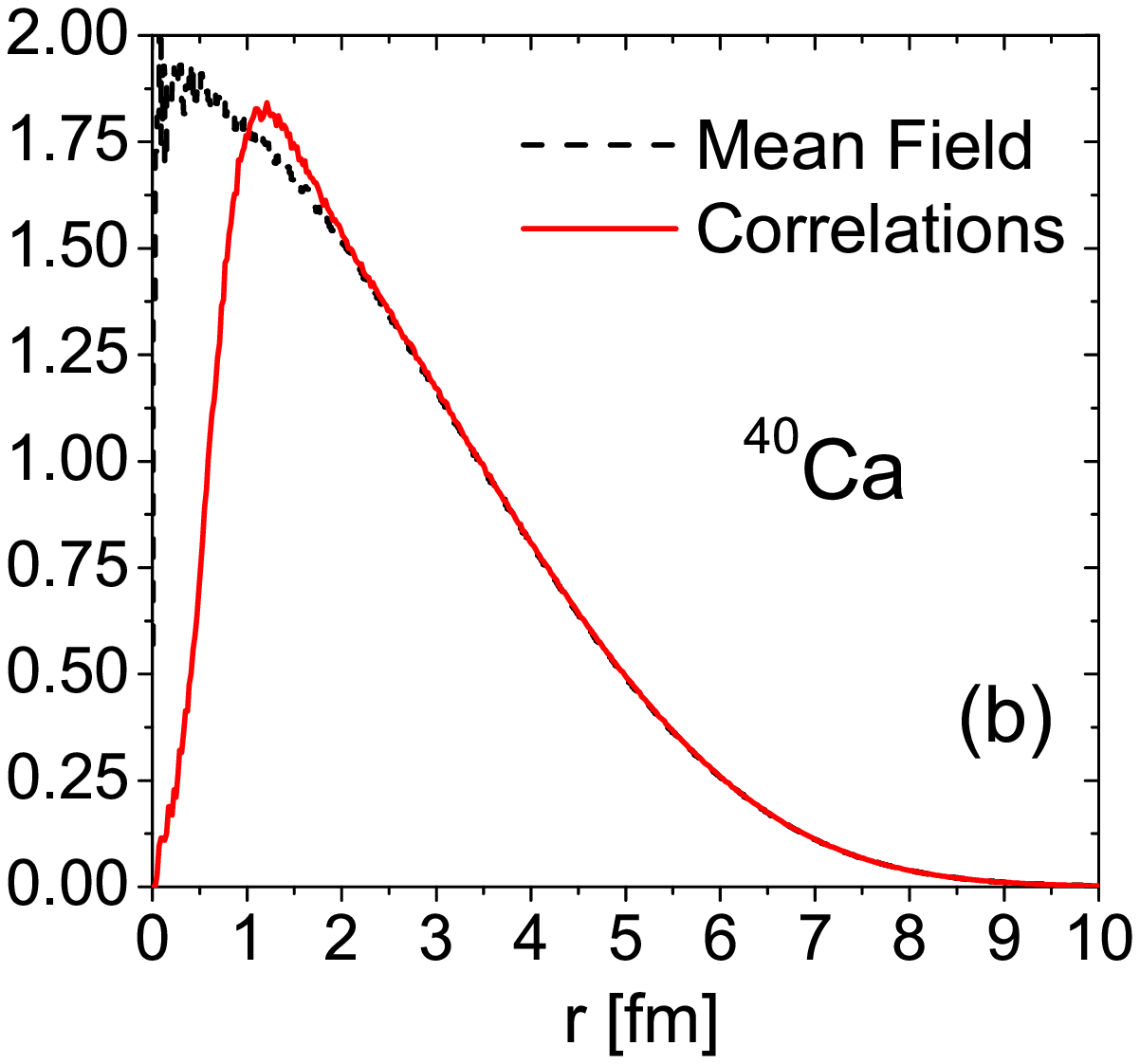}
\hspace{-1.0cm}\includegraphics[height=6.5cm]{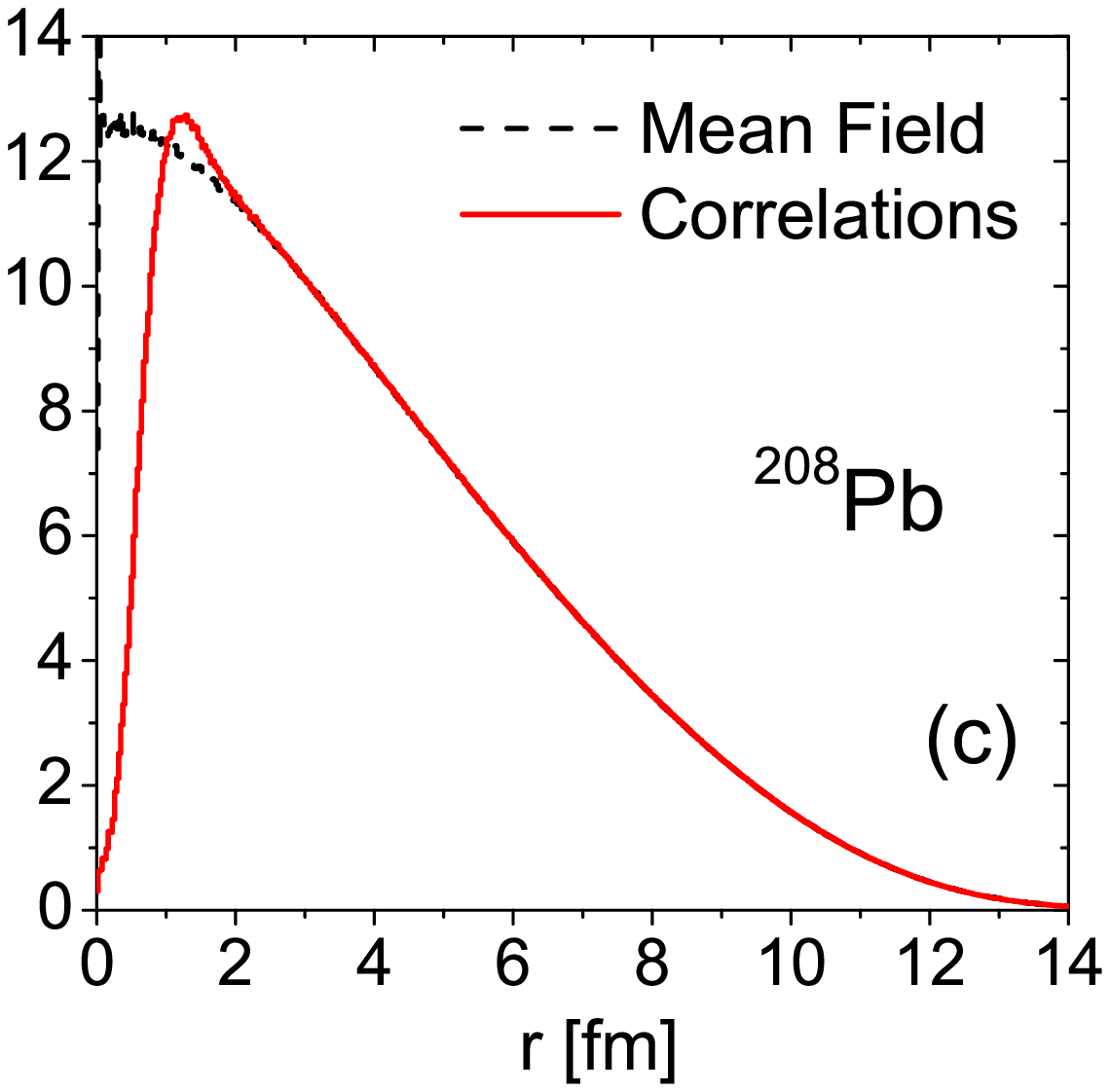}}
\caption{(Color online) The two-body density matrix defined in Eq. (9) calculated for \textit{(a)}
  $^{16}O$, \textit{(b)} $^{40}Ca$, and \textit{(c)} $^{208}Pb$ within our MC code.
  The comparison of the uncorrelated results (\textit{dashes}) and the full correlation results
  (\textit{full lines}) for the first operator $O^{(n=1)}_{12}$ is shown in the various panels.
  The information on state-dependent correlations is contained in the configurations used in the
  calculation, which were produced using the correlation functions of Ref. \cite{Alvioli:2005cz}.
  The curves are normalized to $4\pi\int r^2 dr \rho^{(2)}(r)=A(A-1)/2$.
  }\label{Fig3}
\end{figure*}
It is due to a large contribution of the emission  
from the inner surface generated by removal of a fraction of nucleons due to the collision, which is 
strongly dependent on the geometry of the process. We expect an azimuthal asymmetry which can be exploited 
to determine the centrality of a given collision event and also resolve the sign ambiguity of the impact 
parameter vector.

In Sec. \ref{sec:abl} we compare our results with those of the previous models. Of particular interest
is the analysis of Ref. \cite{Scheidenberger:2004xq}, in which the abrasion-ablation model is used to
define the participant-spectator mechanism and the subsequent spectator system decay and to perform
the estimate of the excitation energy per spectator nucleon. We find that our results for the average
characteristics of the nucleon emission are close to the result of Ref. \cite{Scheidenberger:2004xq}
where  certain inputs from the data were used.
\section{General definitions and method}\label{sec:mc}

The inclusion of central correlations in nuclear configurations can be achieved within a Monte
Carlo Metropolis method by using as a probability function the square of the wave function of
the system taken in the following form:
\beq
\label{defpsi}
\psi_0(\Vec{x}_1,...,\Vec{x}_A)\,=\,\hat{F}(\Vec{x}_1,...,\Vec{x}_A)\,\phi_0(\Vec{x}_1,...,\Vec{x}_A)\,,
\eeq
where $\phi_0(\Vec{x}_1,...,\Vec{x}_A)$ is a Slater determinant of single-particle
densities, the vector $\Vec{x}_i=\left\{\Vec{r}_i;\sigma_i;\tau_i\right\}$ contains the
spatial, spin, and isospin degrees of freedom, and the correlation operator $\hat{F}$ can, in
principle, include the dependence on all the state-dependent operators contained in realistic
nucleon-nucleon potentials; in Ref. \cite{Alvioli:2009ab} a simple product of central correlation
functions was considered, $\hat{F} \equiv F = \prod^A_{i<j}f(r_{ij})$. The calculation which
includes the full product $\hat{F} = \prod^A_{i<j}\hat{f}_{ij}$ with $\hat{f}_{ij}$ state-dependent
correlation functions, is a formidable task; instead, one can consider an expansion of that
product whose first term contains correlation links of the particle under investigation in
the Metropolis search with a second particle only, disregarding correlation links between
the second particle and the others.
The subsequent terms which include such third- and higher-order correlations, namely
three-body clusters linked by spin and isospin dependent two-body correlations, as well
as genuine three-body correlations, will be neglected in the present work. At the same
time the central correlations will be retained to all orders.
This procedure is dictated by the enormous computing power needed for a higher-order
calculation and justified by the fact that realistic calculations based on the cluster
expansion technique show that higher-order calculations, essential for the accurate
determination of quantities such as binding energy and momentum distributions, provides
corrections to the bulk properties of the quantities of relevance, namely diagonal one-
and radial two-body densities, which are small as compared to the accuracy needed by
our study. Instead, we will make use of realistic results obtained within the cluster
expansion method (\cite{Alvioli:2005cz,Alvioli:2007zz}) when applicable.
The one-body densities obtained with our configurations for $A=$16, 40 and 208 are  shown
in Fig.~\ref{Fig2}(a). They coincide with very high precision with the analytic function
used as an input. In Fig.~\ref{Fig2}(b) we present the pair distribution function
\beq
\label{eqpair}
C(r)\,=\,1\,-\,\rho^{(2)}_C(r)/ \rho^{(2)}_U(r)\,,
\eeq
with $\rho^{(2)}_C$ and $\rho^{(2)}_U$ being the correlated and uncorrelated two-body
radial densities, respectively.  
The procedure of Ref. \cite{Alvioli:2009ab} with improved configurations 
\begin{figure}[!htp]
\vskip -0.2cm
\centerline{\includegraphics[height=7.0cm]{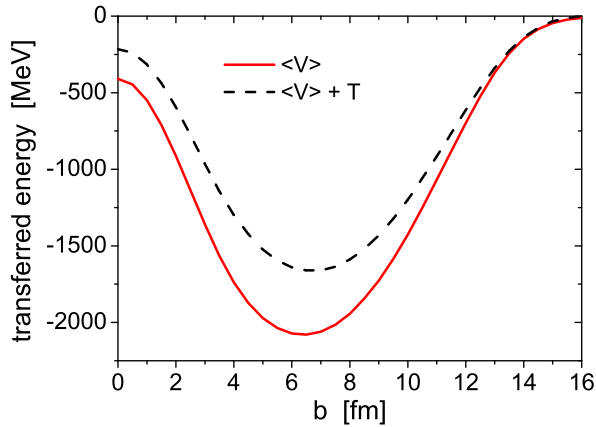}}
\caption{(Color online) Transferred energy in $Pb-Pb$ collisions at $P_{Lab}=160$ GeV,
  calculated within our model at different impact parameters.
  (Solid line) Potential energy due to removing of nucleons
  with no other effects taken into account;
  (dashed line) transferred energy after (i) subtraction of kinetic energy
  of emitted high-momentum nucleons and soft nucleons from the interaction surface
  and (ii) extra energy gained from nucleons absorbed the spectator system.
  }\label{Fig4}
\end{figure}
described above leads
to a better agreement with the realistic  variational calculation from Ref. \cite{Alvioli:2005cz}
than the original procedure which included only central correlations. For comparison we also the
results of Ref. \cite{Alvioli:2009ab} obtained in MC with step ($\theta$ function) and Gaussian
nucleon-nucleon correlation functions which strongly deviate from the realistic calculation
already for $r\sim$ 1 fm.

It is possible to describe the bulk features of the geometry of the the nucleus-nucleus
collisions using the Glauber multiple-scattering model. Within this framework, nucleons
are frozen in their positions during the interaction, which is supposed to be instantaneous.
For a given impact parameter of the colliding nuclei, the impact parameter
$\Vec{b}_{ij}=\Vec{b}_i-\Vec{b}_j$ of the $i$-th projectile and $j$-th target nucleons
are considered, and their inelastic interaction is evaluated on the event-by-event basis
using the impact parameter representation for $NN$ collisions which
leads to the probability of the inelastic nucleon - nucleon at a relative impact parameter $b$:
\beq
\label{pinb}
P_{in}(b)\,=\,1\,-\,\left(1\,-\,\Gamma(\Vec{b}_i-\Vec{b}_j)\right)^2\,,
\eeq
where
$\Gamma(b_{ij}) = \sigma^{tot}_{NN}\,exp(-b^2_{ij}/2B)\,/4\pi B$ is the usual nucleon-nucleon
elastic profile function. In most of our numerical studies reported below we used the parameters
$\sigma^{tot}_{NN}=39$ $mb$ and $B=13.59$ $GeV^2$, corresponding to NA49 energy. The main effect
we neglect here is inelastic shadowing corrections which arise in the Gribov-Glauber approximation
 - these effects primarily affect interactions far from the interaction surface. We also give results
for the energies of RHIC and the Large Hadron Collider (LHC). For illustration purposes, we show in 
\begin{figure}[!htp]
\vskip -0.2cm
\centerline{\includegraphics[height=7.0cm]{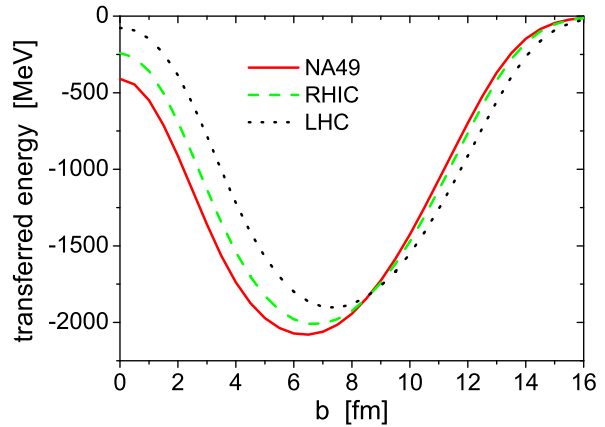}}
\caption{(Color online) Transferred energy in $Pb-Pb$ collisions due to the only
  potential energy transfer. The values of the Glauber parameters correspond to
  NA49 (Solid line), RHIC (dashed line) and LHC (dotted line) energies.
  }\label{Fig5}
\end{figure}
Fig.~\ref{Fig1} the spectator
nucleons after a $Pb-Pb$ collision, for two particular projectile-target configurations. The figure
shows, in addition to protons and neutrons (respectively black and red online), nucleons which before
the event were correlated with an interacting nucleon. These are shown explicitly in white.
The correlated
pairs were identified by checking their relative distance and choosing correlated nucleons on the
basis of the potential energy they can gain due to removal of the neighboring nucleons, as described
in the next sections.
\section{Potential energy calculation}\label{sec:pot}

The first aim of our analysis is to evaluate the energy transferred to the spectator system in the
collision of two heavy nuclei (this calculation does not depend on the details of the fragmentation
discussed in the following sections and could be used in any other models of the high energy nuclear
fragmentation). To this end, we considered the method of the cluster expansion of Refs.
\cite{Alvioli:2005cz} and \cite{Alvioli:2007zz} and found that (i) the inclusion of correlations in the
calculation of the potential energy brings the fraction of the  total potential energy $<V>_{NN}$ due
to $pn$ pairs to about $85\%$ (with the main contribution  due to the existence of tensor interactions
in the nucleon-nucleon potential) and (ii) the state-dependent radial two-body densities $\rho^{(2)}_n(r)$
obtained in Refs. \cite{Alvioli:2005cz} and \cite{Alvioli:2007zz} can be incorporated into our Monte Carlo
code to calculate the total energy transferred to the spectator system in a particular collision and
to determine the fraction of the energy transfer due to the SRCs.

\begin{figure}[!htp]
\vskip -0.2cm
\centerline{
  \includegraphics[height=7.0cm]{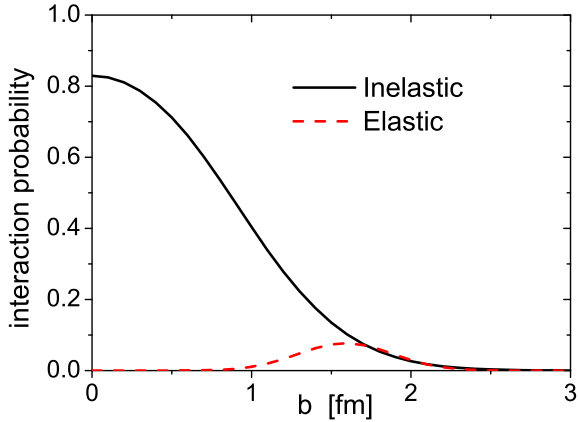}}
\caption{(Color online) Comparison between the probability of inelastic interaction
  $P_{in}(b)$ (\textit{solid line}) of Eq. (\ref{pinb}) and the elastic one $P_{el}(b)$
  (\textit{dashed line}), shown in Eq. (\ref{pelb}), for values of the Glauber
  parameters corresponding to the experiment of Ref. \cite{Appelshauser:1998tt}.
  }\label{Fig6}
\end{figure}
The potential energy contribution to the ground state energy can be calculated according to
\beq
\label{eq1}
\langle V\rangle=
\frac{A(A-1)}{2}\,\sum^6_{n=1}\,\int d\Vec{R}\,d\Vec{r}
\,\rho^{(2)}_n\left(\Vec{R},\Vec{r}\right)
\,v^{(n)}(r)\,,
\eeq
where $\Vec{R}=(\Vec{r}_1+\Vec{r}_2)/2$, $\Vec{r}=\Vec{r}_1-\Vec{r}_2$, and
$\rho^{(2)}_n(\Vec{r}_1,\Vec{r}_2)$ is the state-dependent two-body density
matrix defined as
\beq
\label{eq3}
\rho^{(2)}_n(\Vec{r}_1,\Vec{r}_2)=\int
\sum \prod^A_{j=3}d\Vec{r}_j
\,\psi^\star(\Vec{x}_1,...,\Vec{x}_A)\,\hat{O}^{(n)}_{12}
\psi(\Vec{x}_1,...,\Vec{x}_A)\,.
\eeq
Here the sum extends over all the discrete degrees of freedom so the final result
is a spin-isospin-averaged quantity. Equation (\ref{eq3}) was evaluated within the cluster
expansion method \cite{{Alvioli:2005cz}}. $\hat{O}^{(n)}_{12}$ are the operators
\beq
\label{eq4}
\hat{O}^{(n)}_{12}\,\in\,\left\{\hat{1},\Vec{\sigma}_1\cdot\Vec{\sigma}_2,
\hat{S}_{12}\right\}\,\otimes\,\left\{\hat{1},\Vec{\tau}_1\cdot\Vec{\tau}_2\right\}\,,
\eeq
acting between particles $1$ and $2$,
which inlcude spin- and isospin-dependent nucleon-nucleon potential and corresponding
operators in the nucleus ground-state wave function.
Hence Eq. (\ref{defpsi}) can be rewritten as
\beqy
\label{corfs}
\psi_0(\Vec{x}_1,...,\Vec{x}_A)&=&\prod^A_{i<j}\hat{f}_{ij}
\,\phi_0(\Vec{x}_1,...,\Vec{x}_A)\,=\nn\\
&&\hspace{-2cm}=\prod^A_{i<j}\,\sum^6_{n=1}\,f^{(n)}(r_{ij})\,\hat{O}^{(n)}_{ij}
\,\phi_0(\Vec{x}_1,...,\Vec{x}_A)\,.
\eeqy
The two-body density appearing in Eq. (\ref{eq1}) can be easily separated, within
the cluster expansion as well as in the MC calculation, into the contributions
due to proton-proton ($pp$), proton-neutron ($pn$) and neutron-neutron ($nn$) pairs:
\beqy
\rho^{(2)}_n(\Vec{r}_1,\Vec{r}_2)&=&\nonumber\\
&&\hspace{-2.2cm}=\,\rho^{(2,n)}_{pp}(\Vec{r}_1,\Vec{r}_2)\,+
\,\rho^{(2,n)}_{pn}(\Vec{r}_1,\Vec{r}_2)\,+\,\rho^{(2,n)}_{nn}(\Vec{r}_1,\Vec{r}_2)\,.
\eeqy
The potential energy can thus be written as a sum of three contributions corresponding to
$pp$, $nn$, and $pn$ contributions ($pp$ and $nn$ contributions are identical since the same
single particle orbitals have been used both for protons and neutrons states). Calculations
show that for \textit{central} correlations (the total result in this case has a wrong absolute
value due to the lack of realistic correlations needed to approximate the experimental values
of binding energies), the individual contributions for $pp$ and $pn$ pairs are exactly proportional
to the fraction of $pp$ pairs, (23\%), and $pn$ pairs, (53\%), respectively. At the same time,
if the full correlations are included, this proportionality no longer holds: the $pp$ contribution
represents 8\% of the total, and the $pn$ the 83\% of the total; if we translate
this to the contributions of isospin $0$ and $1$ pairs, we obtain a ratio of 74\% for $I=0$ and
26\% for $I=1$. These features of the short-range nuclear structure can be exploited within the
formalism we are going to describe.

We now outline how to incorporate the information presented in this section in our MC code.
The realistic calculation of Ref. \cite{Alvioli:2005cz} provides us with the contribution to
the total potential energy of a given pair of nucleons,
while the corresponding calculation of total kinetic energy from momentum distributions
for medium-weight nuclei, in Ref. \cite{Alvioli:2005cz}, and for heavy nuclei, in Ref.
\cite{CiofidegliAtti:1995qe}, can be used to evaluate the total excitation energy of the
nucleons which do not experience inelastic interactions.

For each $AA$ event, we select spectator and interacting nucleons in both nuclei, using
Eq. (\ref{pinb}) as the interaction probability; this quantity depends on the (nucleon
pair) relative impact parameter $b_{ij}$. Accordingly the probability \textit{not to interact}
is given by
\beq
P^{survival}_i(b_i) = \prod^A_{j=1} (1- P(b_i-b_j))
\eeq
Hence we can calculate for a given event the  potential energy which is freed by instantaneous
removal of the inelastically interacting nucleons. It represents the amount of the energy available
for freeing nucleons from the  bound state and generating their kinetic energy (in a sense our
approximation resembles the Koltun sum rule \cite{Koltun:1974zz} for removal of one nucleon from
the nucleus). The total potential energy of the nucleus can be calculated in terms of the two-body
density of the system or, more specifically, from the radial two body-density defined as follows:
\beq
\label{eq2}
\rho^{(2)}_n(|\Vec{r}|)=\hspace{-0.1cm}\int d\Vec{R}\,\rho^{(2)}_n
\hspace{-0.1cm}\left(\Vec{r}_1=
\Vec{R}+\frac{\Vec{r}}{2},\Vec{r}_2=\Vec{R}-\frac{\Vec{r}}{2}\right)\,.
\eeq
\begin{figure*}[!htp]
\centerline{\includegraphics[width=9.0cm]{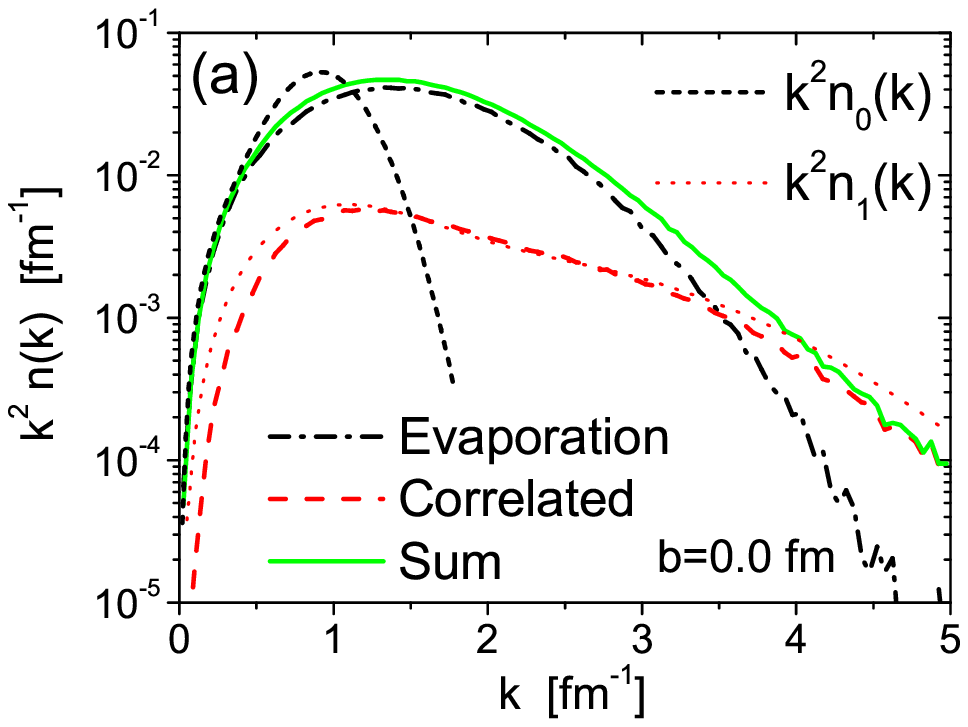}
\hspace{0.0cm}\includegraphics[width=9.0cm]{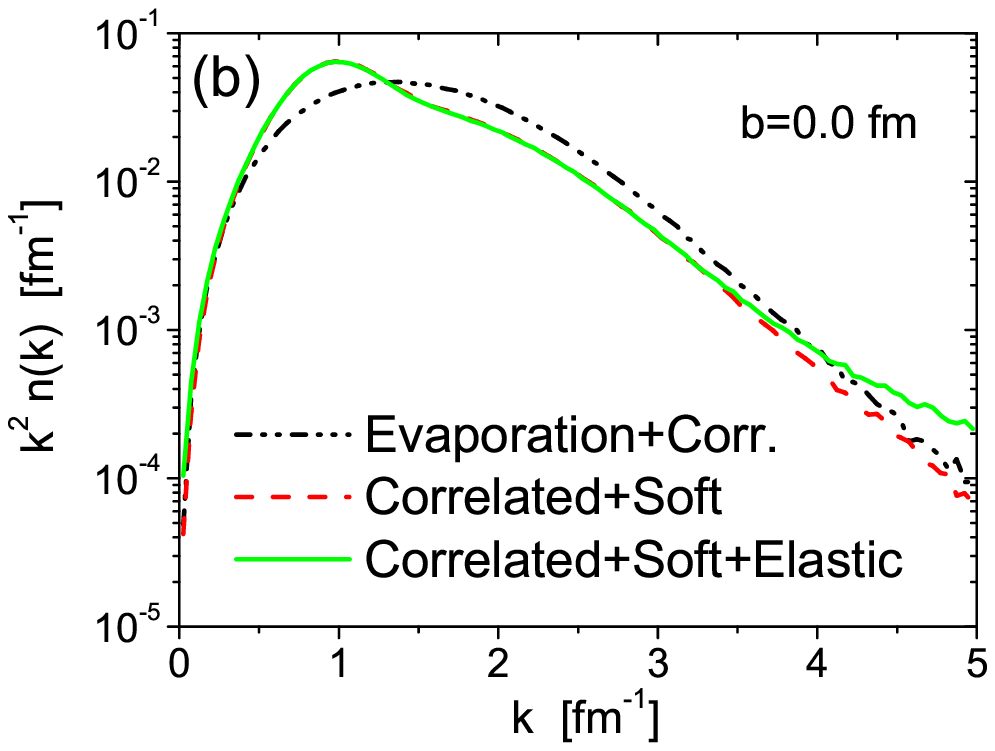}}
\caption{(Color online) The momentum distributions $k^2\,n(k)$ of directed flow of emitted
  nucleons in $Pb-Pb$ collisions at $P_{Lab}=160$ GeV, calculated within our model at $b=0$
  within different approximations. \textit{(a)} Comparison of the input momentum distributions
  $n_0(k)$ and $n_1(k)$ against the MC output considering only \textit{evaporation} and
  emission of \textit{correlated} nucleons (no primary elastic scattering and no additional
  soft nucleons emitted from the surface taken into account); the contribution of the high-momentum
  tail of $n_1(k)$ is evident in the \textit{sum}.
  \textit{(b)} The total momentum distributions from MC within the various approximations
  described in the text. The normalization of each curve is set to the actual number $N$ of
  nucleons falling into the corresponding definition, according to $4\pi\int dk k^2n(k)=N$.
  }\label{Fig7}
\end{figure*}
This quantity may be easily calculated in the MC approach by considering all the pairs and building
the distribution of their relative distances. It is then straightforward to calculate the fractions
of the total two-body  potential energy which are due to the interaction between two spectators, two
nucleons which experienced inelastic collisions, and mixed pairs. 
Comparing these fractions with the realistic
calculations, in (small) intervals in $r$ corresponding to the bins used in the Monte Carlo
determination of the densities, properly scaling the MC densities to the ones obtained in the
realistic calculations and rescaling to the experimental value of the binding energy of the 
initial nucleus, we can define
the fractions $<V>^{SPE}_{NN}$, $<V>^{INT}_{NN}$ and $<V>^{MIX}_{NN}$ accordingly. 
The MC procedure is not accurate enough as far as the state-dependent
radial two body densities $\rho^{(2)}_n$ are concerned for a meaningful determination of the
expectation value of the realistic, state dependent potential
$\hat{V}_{ij}=\sum_n v^{(n)}(r_{ij}) \hat{O}^{(n)}_{ij}$ (such a calculation would require a much
more accurate balance of several positive and negative parts which is currently practical) 
The procedure
can also be applied to the individual $pn$ and $pp$ pairs. The two-body densities obtained from
the generated nuclear configurations are presented in Fig.~\ref{Fig3} for different nuclei. The
figure shows both the results corresponding to uncorrelated and correlated configurations for the
radial two-body density associated with the central operator,  namely $\rho^{(2)}_{n=1}$ in Eqs.
(\ref{eq3}) and (\ref{eq2}).

The outlined procedure determines the fraction of potential energy due to mixed pairs of nucleons
on the event-by-event basis. The results of the calculation of this quantity (which determines the
total excitation energy of the spectator system which is released in the process of nucleon / fragment
emission) are presented in Fig.~\ref{Fig4} for the case of collisions at the momentum of 160 GeV/$c$
per nucleon for which the most extensive high energy studies of the fragmentation were performed.
We plot the potential energy due to the disrupted pairs and the same (negative) quantity plus the
(positive) kinetic energy due to the emission of high-momentum nucleons. Fig.~\ref{Fig4} also shows
the dependence of the transferred energy as a function of the impact parameter. Also, we show in
Fig.~\ref{Fig5} the dependence of the transferred energy on the value of the cross section of $NN$
interactions taken for values corresponding to NA49, RHIC and LHC energies. One can see that for
average impact parameters the energy dependence is rather weak. At the same time, for small impact
parameters, the energy transfer drops with an increase of energy since the chances for spectators to
survive decrease due to increase of $\sigma^{tot}_{NN}$, while the increase of $\sigma^{tot}_{NN}$
leads to increase of the probability to interact with several nucleons at large impact parameters,
leading to the increase of the energy transfer at large $b \ge 10$ fm.

The details of our model for the description of emission of nucleons are given in the next sections.
\begin{figure*}[!htp]
\vskip -0.6cm
\centerline{\includegraphics[width=20.0cm]{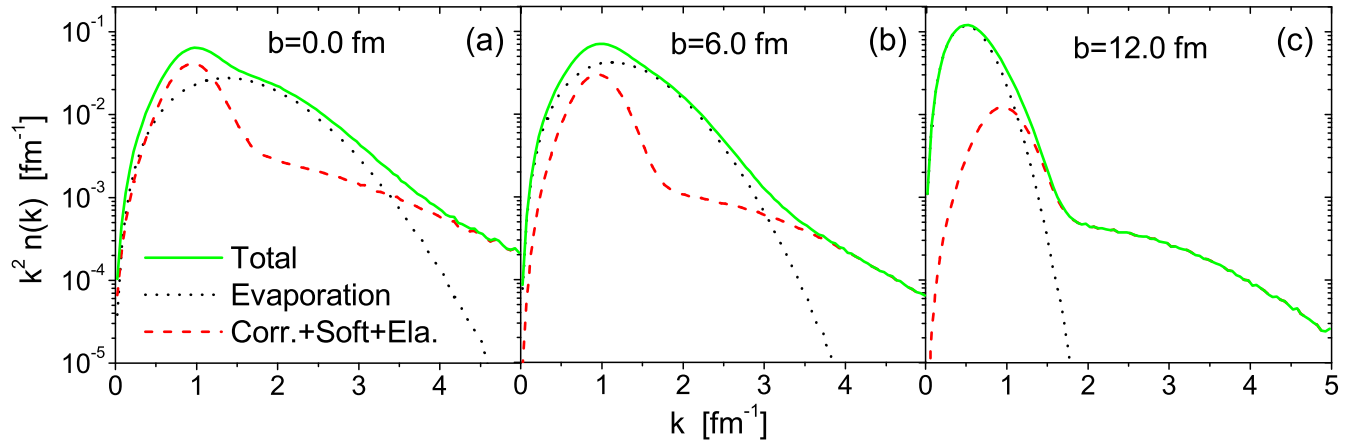}}
\vskip -0.6cm
\caption{(Color online) The momentum distributions $k^2\,n(k)$ of emitted nucleons in
  $Pb-Pb$ collisions at $P_{Lab}=160$ GeV, calculated within our full model. We show
  the different contributions due to evaporation (\textit{black dotted}) and the sum of all
  other nucleon emission mechanisms, namely soft nucleons from the surface, correlated nucleons
  and elastically scattered nucleons (\textit{red dashed}); the total momentum distribution
  is also shown (\textit{green solid}).
  [(a)-(c)] The momentum distributions plotted for
  impact parameters of $b=0$, $b=6$, and $b=12$ fm, respectively. Note that for central collisions
  the most relevant contribution to the total normalization is due to the high-momentum part, as
  expected, while  when $b$ increases the dominant contribution is due to evaporation.
  }\label{Fig8}
\end{figure*}
\section{Nucleon emission and momentum distributions}\label{sec:momdis}

We have developed a model for nucleon emission as a function of the solid angle
$d\Omega = d(\cos\theta)d\phi$ from one of the colliding nuclei in a Pb-Pb collision.
The model takes into account several different mechanisms for the nucleon emission reflecting
the geometry of the high energy $AA$ interaction. Previously only evaporation of nucleons
from the excited spectator system resulting from the collision was considered and the total
energy released in the emission was treated as an input parameter.
In the previous section we developed a  novel approach  which allows us to calculate the total excitation
energy on an event-by-event basis. Part of this energy is emitted in a surface process resulting from
the sudden removal of a part of the nucleus with large energies transferred locally to nucleons which
had strong bonds with removed nucleons. We first consider emission of the high momentum
(Sec. \ref{sec:corr})  and low momentum nucleons (Sec. \ref{sec:surf}) from the surface.
We also consider nucleons emitted in elastic interactions of two nucleons of the colliding nuclei which
gives an important contribution for certain emission angles (Sec. \ref{sec:primela}). The nucleons
generated in the surface emission and elastic scattering propagate with a large fraction of the events
through the spectator system and transfer energy to the bulk of the spectator system (Sec. \ref{sec:ela}).
Taking into account all these effects we evaluate the total energy available for evaporation of soft
nucleons (Sec. \ref{sec:evap}).

We discuss in detail the different mechanisms and how we calculate the  momentum distributions of emitted
nucleons in the following subsections.
\subsection{Correlated nucleons emission}\label{sec:corr}

For a given impact parameter $b$, we calculate the interaction between individual nucleons within the
Glauber multiple scattering model as described in Sec. \ref{sec:intro} and classify the different
pairs of the nucleons  as described in Sec. \ref{sec:pot}. Next, we consider the \textit{mixed} pairs,
\textit{i.e.} those pairs in which one of the nucleons is a spectator and the second is an interacting
one, and define nucleons as \textit{active} on the basis of the potential energy transferred to them.
These nucleons should be the candidates among which we choose actually \textit{correlated} nucleons.
We have combined information about the short-range character of correlations and the relatively large
contribution of correlations to the total potential and kinetic energies. To this end, we define as
active the spectators which are, in a given event, at a distance below some $r_{max}$ from at least
one interacting nucleon. We estimate $r_{max}$ from the following considerations.
The SRC originate from the internucleon
distances $\le$1.4 fm, hence $r_{max} > 1.4 fm$. At the same time for $r > $ 2 fm the internucleon
potential corresponds to energies significantly smaller than the Fermi energy. Hence a removal
of such a long range bond is likely result in collective excitations of the residual system.
Hence a reasonable value of this parameter of the model is  $2 fm \le r_{max} \le  1.4 fm$.
\begin{figure*}[!htp]
\vskip -0.2cm
\centerline{\includegraphics[height=7.0cm]{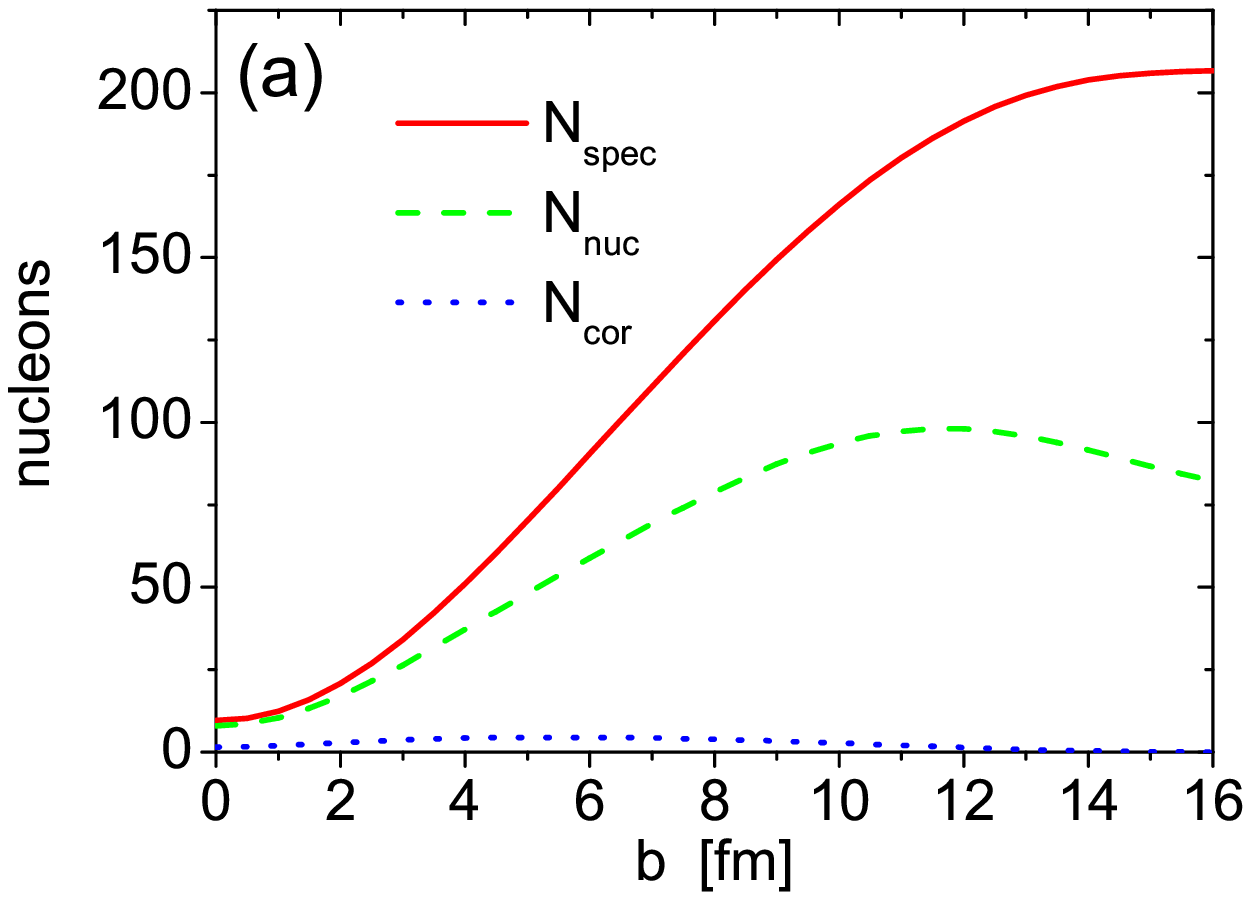}
\hspace{0.0cm}\includegraphics[height=7.cm]{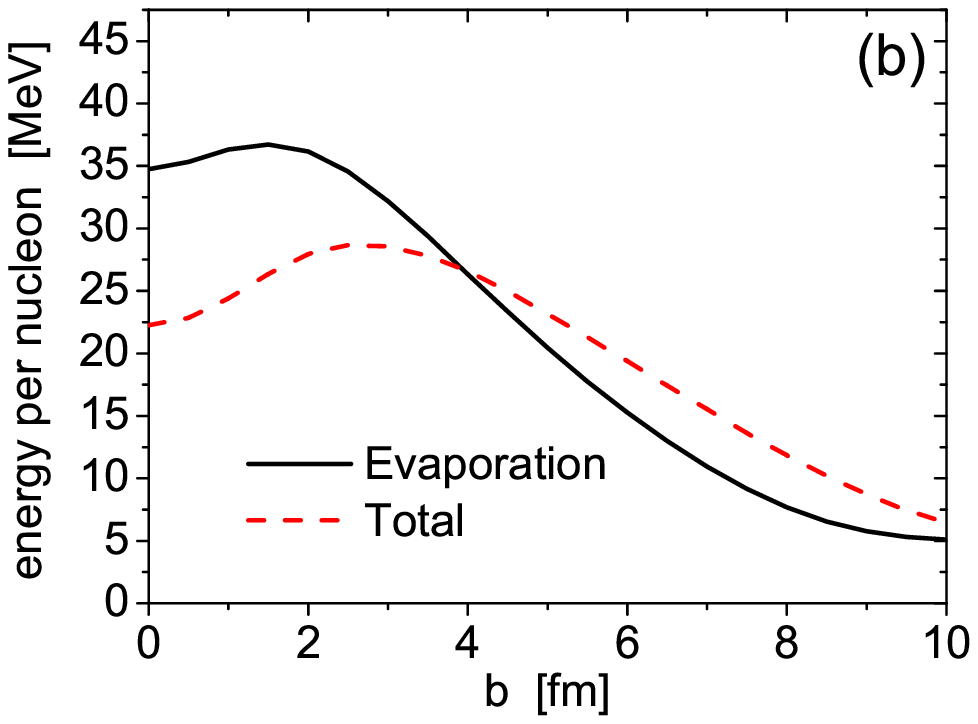}}
\caption{(Color online) \textit{(a)} Spectators in a $Pb-Pb$ collision within a Glauber
  model at NA49 energy ($P_{Lab}=$160 GeV per nucleon). The total number of spectators $N_{spec}$,
  unbound nucleons $N_{nuc}$, and high-momentum nucleons $N_{cor}$ are shown. Data from the
  NA49 experiment \cite{Appelshauser:1998tt} was used to determine the spectators/emitted
  nucleons ratio.
  \textit{(b)} The kinetic energy per emitted soft nucleon as calculated within our model. 
  We show the case in which only evaporation is taken into account
  (black solid line) and the case when all the effects described in Sec. \ref{sec:momdis} 
  are taken into account (red dashed line).
}\label{Fig9}
\end{figure*}
In the following we will use  as a base value  $r_{max}$ 2 fm.
The  sensitivity to the variation of $r_{max}$ is explored in Sec. \ref{sec:mdep}
Next we calculate the energy transferred to each of the active nucleons,
making use of our realistic estimate of the potential energy due to the removal of the interacting
nucleons. After ranking the active nucleons starting from the one to which the highest energy is
transferred, we choose as correlated the first 25\% of them, according to the estimated fraction
of correlations in heavy nucleons \cite{CiofidegliAtti:1995qe}. We choose, among the active ones,
those nucleons which have larger potential energies and at the same time are at smaller distances
from one of their interacting partners, since spatial proximity is a basic requirement for two
nucleons to be correlated. This is also clear from the short-range character of correlation
functions used in many-body calculations.
A delicate feature we need to deal with is that in quantum mechanics there is no one-to-one relation
between the potential $V$ and kinetic energy; the relation $T = -V +\epsilon$, with $\epsilon$ the
binding energy per nucleon, is valid only in average. Hence we will use a probabilistic algorithm
of assigning correlated nucleons random momentum vectors with modulus given by the probability
distribution $n_1(k)$, which is the correlated, high momentum tail of the momentum distribution from
Ref. \cite{CiofidegliAtti:1995qe}. In the model of Ref. \cite{CiofidegliAtti:1995qe} the total momentum
distribution in a nucleus is modeled as a mean field, low momentum part $n_0(k)$ plus a correlated,
high-momentum part $n_1(k)$, which accounts for the 75\% and 25\%, respectively, of the total normalization
in a heavy nucleus.
The procedure will be discussed again in Sec. \ref{sec:mdep}, after several intermediate steps are
described.
Note that the production of nucleons from the correlations has forward-backward
asymmetry which for moderate momenta is given by the flux factor $(1+k_3/m_N)$, where
$k_3$ is the longitudinal component of the nucleon momentum in its rest frame. So more
nucleons are emitted forward (along the beam direction of the projectile) in the rest
frame of a nucleus. In this paper to simplify the discussion we consider quantities
symmetrized over $k_3 \to -k_3$.

At the next step active nucleons propagate through the residual system and could be absorbed.
This effect is evaluated in Sec. \ref{sec:ela}, using fits to the $NN$ elastic scattering data
\cite{Arndt:2007qn}.
\subsection{Direct emission of uncorrelated nucleons from the inner surface}\label{sec:surf}

The number of correlated nucleons determined in Sec. \ref{sec:corr} is only a fraction of active
nucleons whose "broken links" with the removed nucleons produce a large fraction of the available
energy and which are located at $r\le r_{max}$ from at least one of the removed nucleons.
The rest of the nucleons, located near the inner surface resulting from the removal of the
interacting nucleons, must be considered as escaping the system as well, but with a momentum
distribution given by the low-momentum, mean-field part $n_0(k)$ of the model of Ref.
\cite{CiofidegliAtti:1995qe}. This assumption is reasonable since we assume the active nucleons
to be emitted with the same momenta they had in the nucleus, but it is clearly an approximation.
\subsection{Emissions from elastic scattering}\label{sec:primela}

In addition to the spectators which underwent inelastic scattering, there are (primary)
\begin{figure*}[!htp]
\centerline{\includegraphics[width=7.0cm]{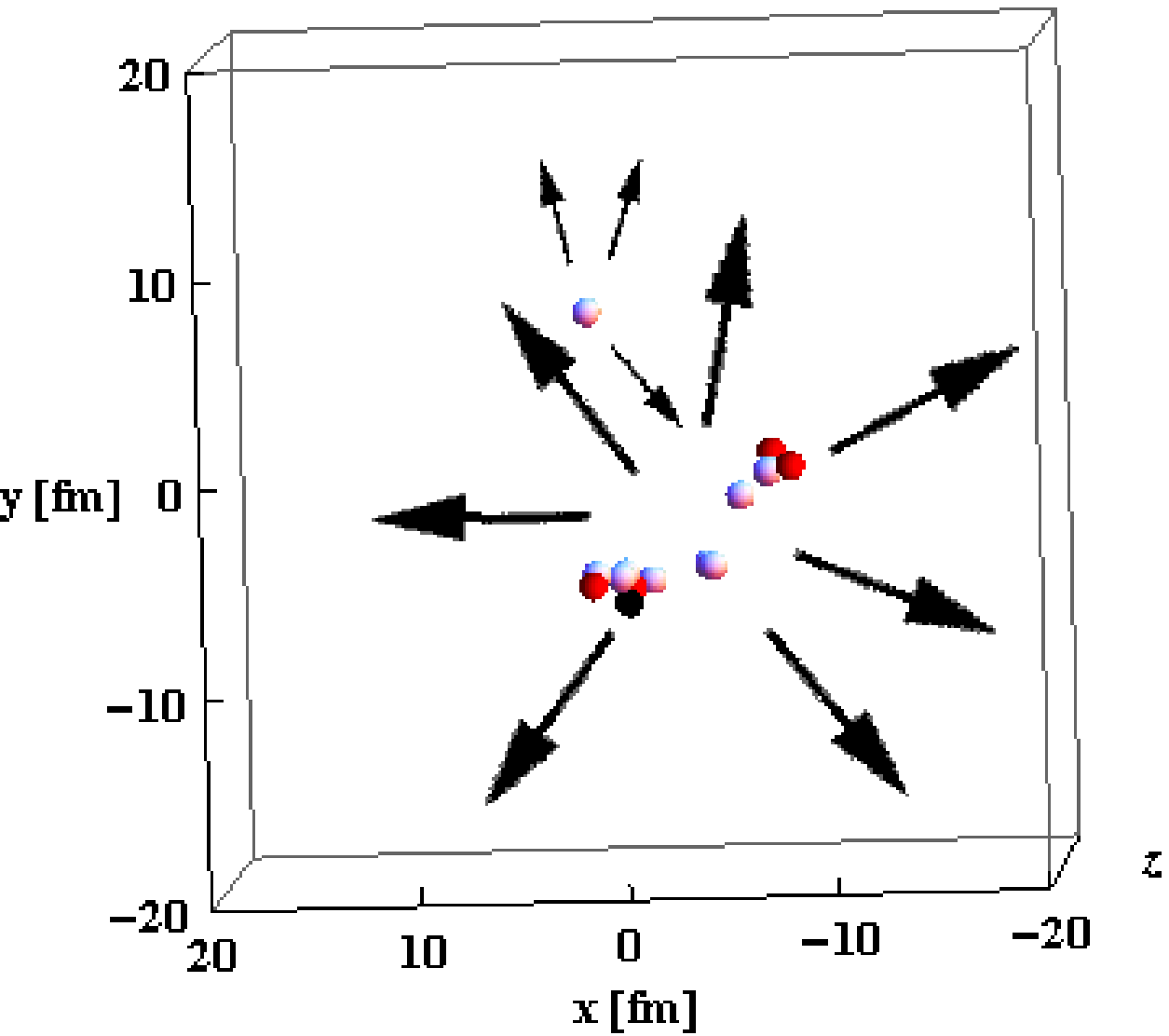}
\hspace{0.2cm}\includegraphics[width=7.0cm]{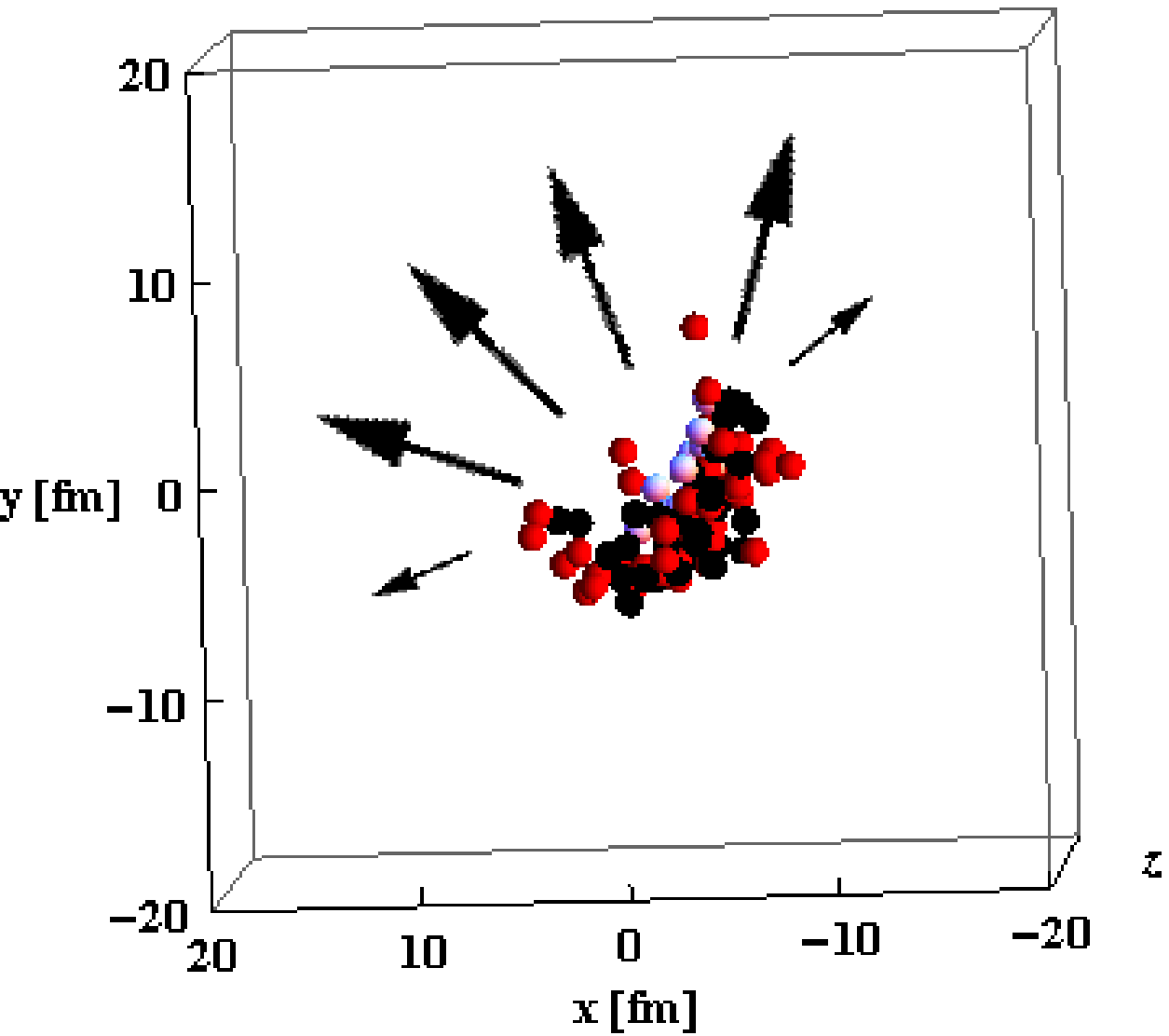}}
\vskip 0.2cm
\centerline{\includegraphics[width=7.0cm]{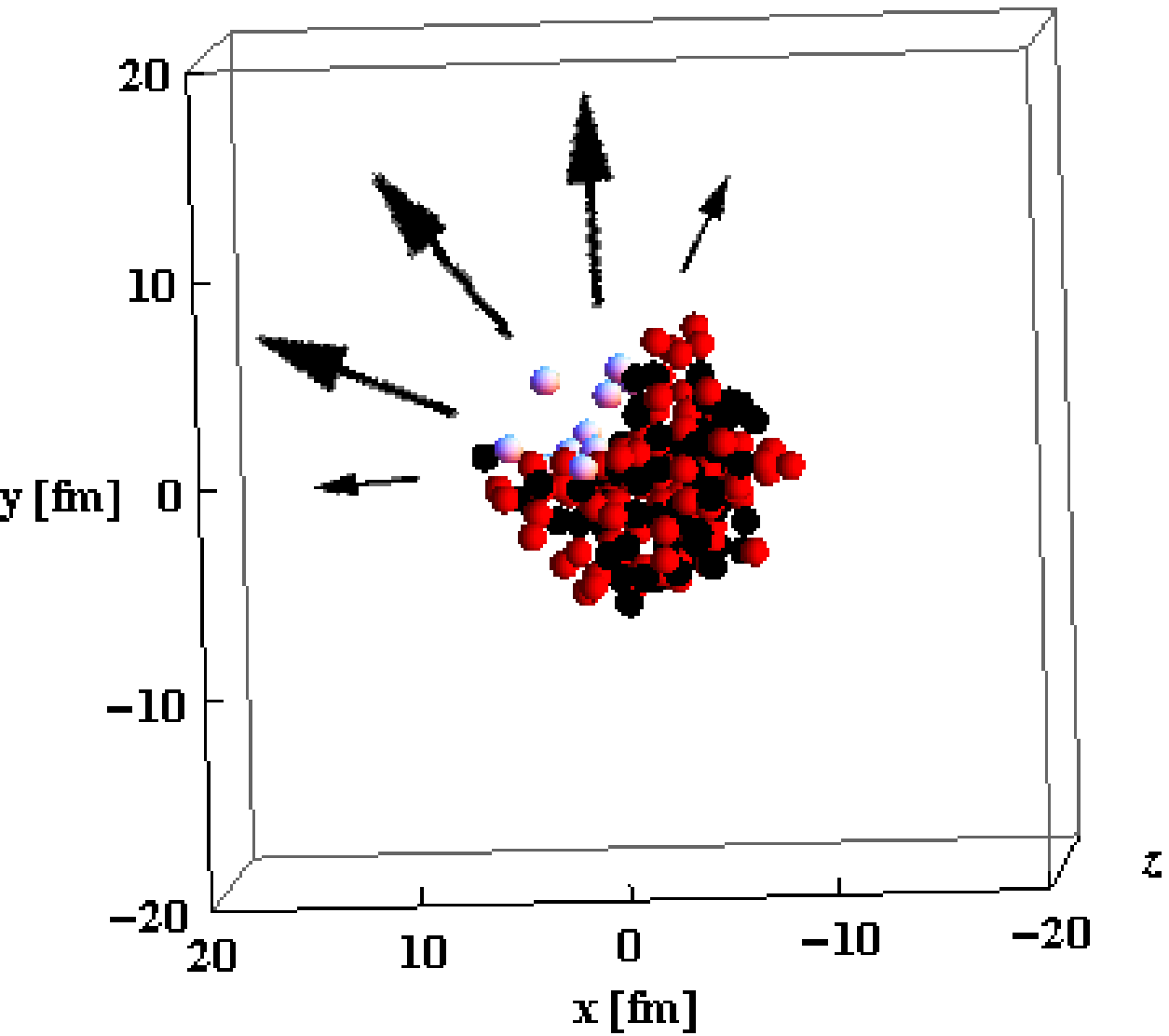}
\hspace{0.2cm}\includegraphics[width=7.0cm]{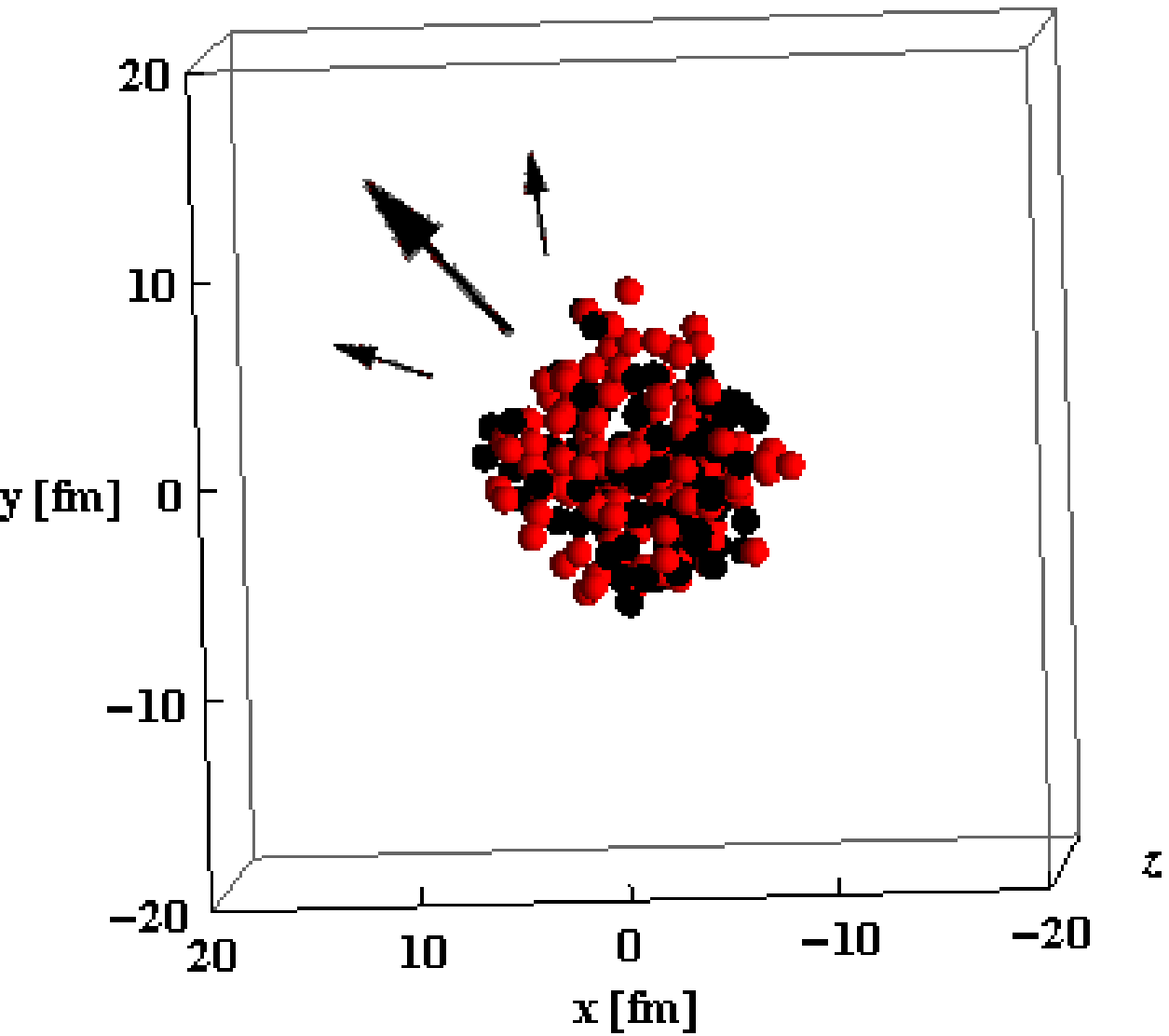}}
\vskip -12.2cm                                   %
\centerline{\large\textbf{(a)}\hspace{6.7cm}\textbf{(b)}\hspace{-4.4cm}}
\vskip 6.0cm                                     %
\centerline{\large\textbf{(c)}\hspace{6.7cm}\textbf{(d)}\hspace{-4.4cm}}
\vskip 5.3cm                                     %
\caption{(Color online) Sketch of the asymmetry of emission of high-momentum, correlated
  nucleons defined in Eq. (\ref{asymdef}) and quantitatively evaluated in Fig.~\ref{Fig12},
  in Pb-Pb collisions; the preferred direction of emission is shown with arrows as a function
  of the impact parameter which is oriented as in Fig.~\ref{Fig1} and its values through the
  different panels are \textit{(a)} $b=$1 fm, \textit{(b)} $b=$5 fm, \textit{(c)} $b=$10 fm
  and \textit{(d)} $b=$15 fm.
}\label{Fig10}
\end{figure*}
nucleons which scattered elastically. We model the probability of high-energy $NN$
elastic scattering $P_{el}(b)$ in the following way. 
It must obey the sum rule
\beq
\int d\Vec{b}\,P_{el}(b)\,=\,\sigma^{el}_{NN}
\eeq
(at NA49 energy we have $\sigma^{el}_{NN}=5.72$ mb), it must be a function of
$b=b_{ij}=|\Vec{b}_i-\Vec{b}_j|$ and vanish when $P_{in}(b)$ does.
A reasonable distribution appears to be
\beq
\label{pelb}
P_{el}(b)\,=\,0.077\,e^{-5.7\,(b-1.59)^2}
\eeq
which is compared with $P_{in}(b)$ of Eq. (\ref{pinb}) in Fig.~\ref{Fig6}.
Using Eq. (\ref{pelb}), we select elastically scattering nucleons and eventually
take into account their recoiling partners as emerging from the event and propagate them
through the spectator medium. We calculate  the momentum and emission angle of the recoiling
nucleon as follows. The target nucleon has momentum $k=\sqrt{k^2_t+k^2_3}$ (with probability
distribution $n_1(k)$ from Ref. \cite{CiofidegliAtti:1995qe}); the scattered nucleon has momentum
$p=\sqrt{p^2_t+p^2_3}$; the four-momentum transfer $q$ can be assumed to have $q_0=q_3$ since
we are at high energies so $t=q^2=q^2_0-q^2_t-q^2_0=-q^2_t$; $t$ can be randomly generated
using the elastic scattering amplitude
\beq
\frac{d\sigma^{el}_{NN}}{dt}\,=\,\frac{{\sigma^{tot}_{NN}}^2}{16 \pi}\,e^{B\,t}\,,
\eeq
using the differential $NN$ elastic cross-section data for the energy corresponding to the energy
of the $AA$ collisions. This determines $q_t$ and $p_t=q_t+k_t$. Since the light-cone fraction, 
$\alpha$,  carried by the nucleon is conserved in the elastic scattering we find
\beq
\alpha\,=\,1\,+\,\frac{k_3}{m}=\,\frac{\sqrt{p^2_t+p^2_3+m^2}\,-\,p_3}{m}
\eeq
leading to
\beq
p_3\,=\,\frac{p^2_t+m^2-m^2 \alpha^2}{2 m \alpha}
\eeq
This allows to determine the scattering angle $\theta$ is $\tan{\theta}\,=\,p_t / p_3$. Once
\begin{figure*}[!htp]
\vskip -0.2cm
\centerline{\includegraphics[height=7.0cm]{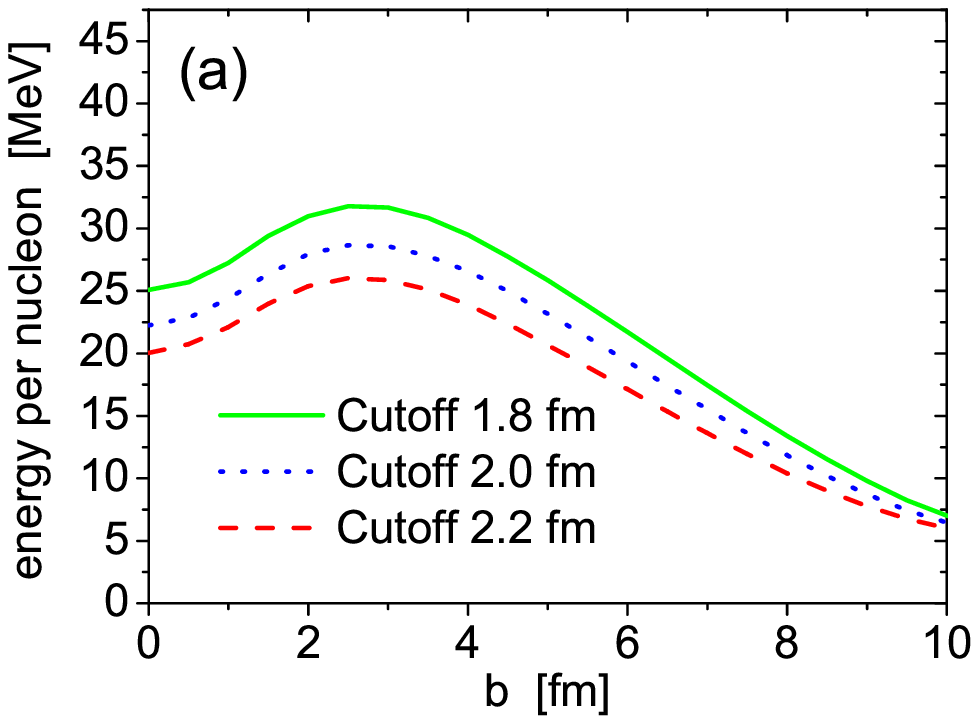}
\hspace{0.0cm}\includegraphics[height=7.cm]{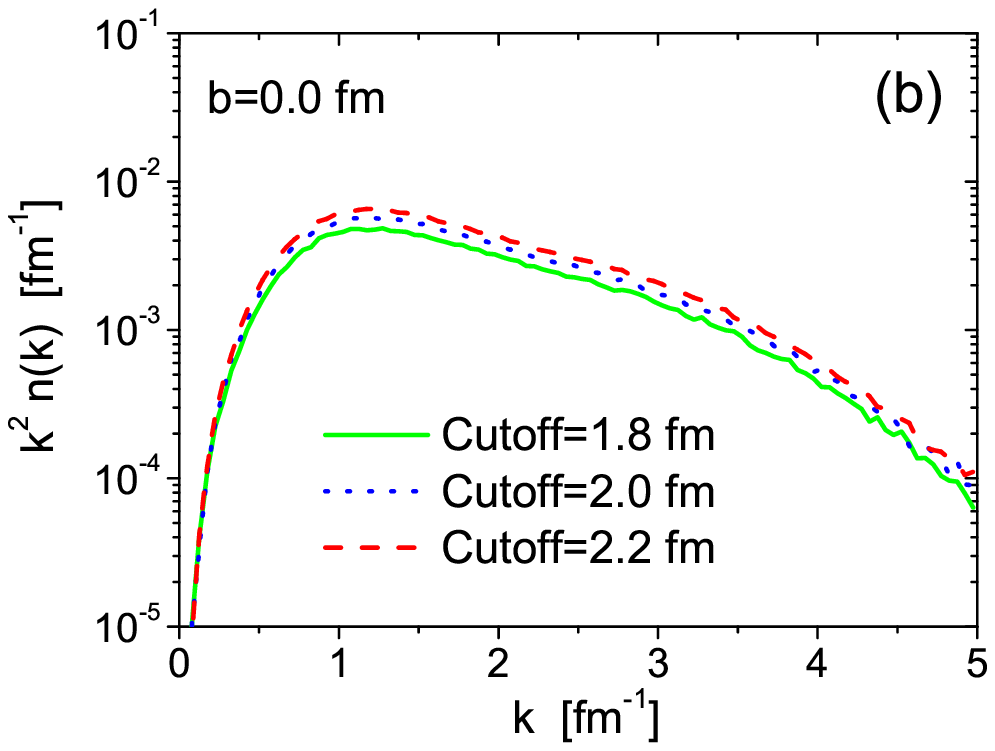}}
\caption{(Color online) Dependence of (a) the total potential energy per evaporating
  nucleon and (b) of the correlated contribution to the momentum distribution
  \textit{(b)}. We show three values of the cutoff distance below which we select
  \textit{active} nucleons among the spectator ones. The distance is calculated from
  the nearest interacting nucleon.
}\label{Fig11}
\end{figure*}
elastically scattered nucleon is generated we can use the method described in \ref{sec:ela} 
to determine whether the generated nucleon is absorbed during the propagation, with its energy 
going into additional heating of the spectator system, or it is emitted with its recoiling
momentum, to be collected in the final momentum distribution. Note that the elastic mechanism
predominantly contributes for $\alpha\sim 1$. The data from Ref. \cite{Anderson:1982jx} do
indicate an extra contribution at the corresponding kinematics.
\subsection{Elastic scattering in the final state}\label{sec:ela}

The nucleons which are emitted in the processes described in the previous subsections may 
be absorbed while propagating through the spectators by thermalization due to elastic
rescattering, leading to the heating of the spectator system. We can evaluate the attenuation
due to the elastic scattering using $d\sigma^{el}_{NN}/dT$ of Ref. \cite{Arndt:2007qn}, given 
as a function of incident nucleon kinetic energy $T=T_{lab}$ and center-of-mass scattering angle
$\theta_{CM}$. We will consider as thermalized those nucleons which, as a consequence of elastic
rescattering, are left with kinetic energy smaller than $T_{min}=\sqrt{k^2_{min}+m^2}-m$ kinetic 
energy, where $k_{min}=250$ MeV/$c$, the typical minimal momentum scale of the SRCs in nuclei.
First, we consider as interacting elastically two nucleons whose transverse separation (with respect
to the direction of propagation of the nucleon under investigation) is
$b_{ij}\,<\,\sqrt{\sigma^{el}_{NN}/\pi}$ with $\sigma^{el}_{NN}$ evaluated at the corresponding
incident momentum, taken from the tabulated values of Ref. \cite{Arndt:2007qn}.
If $k=k_{lab}=\sqrt{(T-m)^2-m^2}$ is the emitted nucleon momentum in the nucleus rest frame hitting 
a spectator nucleon at rest which recoils with kinetic energy $T_R$  we have
\beq
t\,=\,- 2 m T_R\,=\,- 2 p^2 (1-\cos{\theta_{CM}})
\eeq
where $p=k_{CM}=\sqrt{\frac{1}{2}m\sqrt{k^2+m^2}\,-\,\frac{1}{2}m^2}$ is the incident nucleon
momentum in the center of mass. Then, the probability for the nucleon emerging from the collision 
of having more than $T_{min}=T_R-T$ is given by
\beq
\label{pelint1}
\int^T_{40} dT^\prime\,\frac{d\sigma^{el}_{NN}(T^\prime,\theta_{CM})}{dT^\prime}\,/\,
\int^\infty_{40} dT^\prime\,\frac{d\sigma^{el}_{NN}(T^\prime,\theta_{CM})}{dT^\prime}
\eeq
where the lower limit of integration is dictated by the fact that we have from
Ref. \cite{Arndt:2007qn} tables for $40\,\mbox{MeV}\,<\,T\,<\,600\,\mbox{MeV}$
and the upper limit can safely be considered large enough.
Nucleons with  $T<40$ MeV are considered as being absorbed the the spectator medium.
In Eq. (\ref{pelint1}) we use $d\sigma^{el}_{pp}$ both for $pp$ and $nn$ scatterings, 
and $d\sigma^{el}_{pn}$ for $pn$ and $np$.
In the case of absorption by the spectator system, we add their kinetic energy to the
amount of energy available for the subsequent evaporation. The same procedure has been 
used to calculate the propagation of the nucleons generated in the 
\textit{primary (high energy)} 
elastic NN scattering (Sec. \ref{sec:primela}).

It is worth noting here that to the best of our knowledge the contribution of the high energy 
elastic scattering mechanism with or without subsequent absorption was not considered before. 
Of separate interest here is that we calculate this effect (like all other 
quantities defined in this work) as a function of the $AA$ impact parameter $b$.
\begin{figure*}[!htp]
\vskip -0.2cm
\centerline{\includegraphics[height=7.0cm]{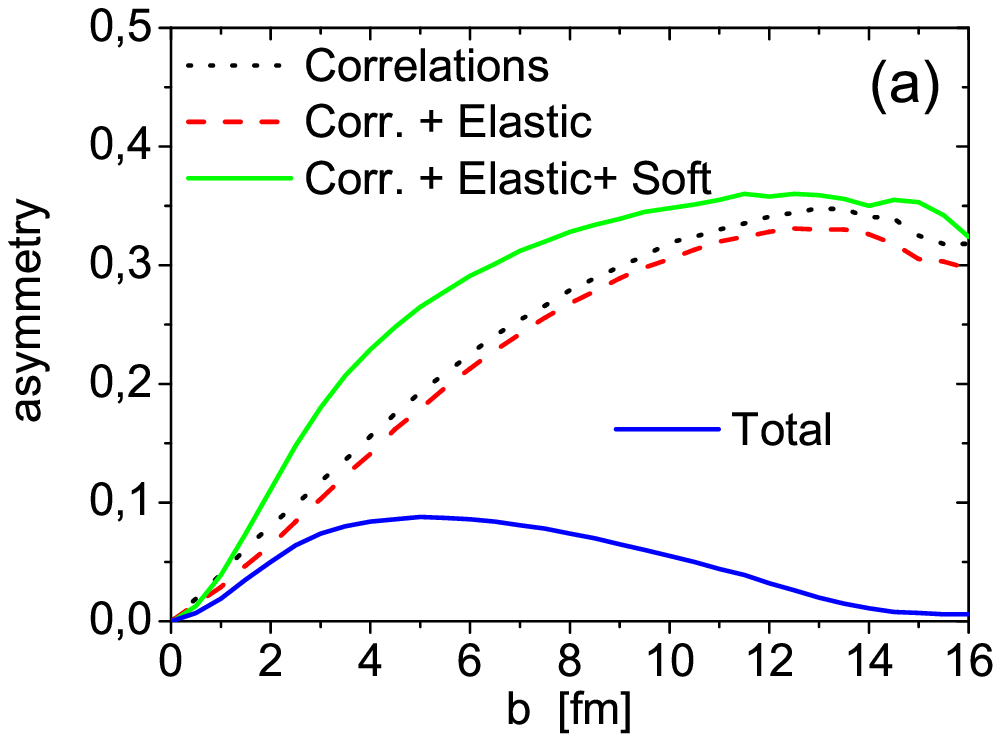}
  \includegraphics[height=7.0cm]{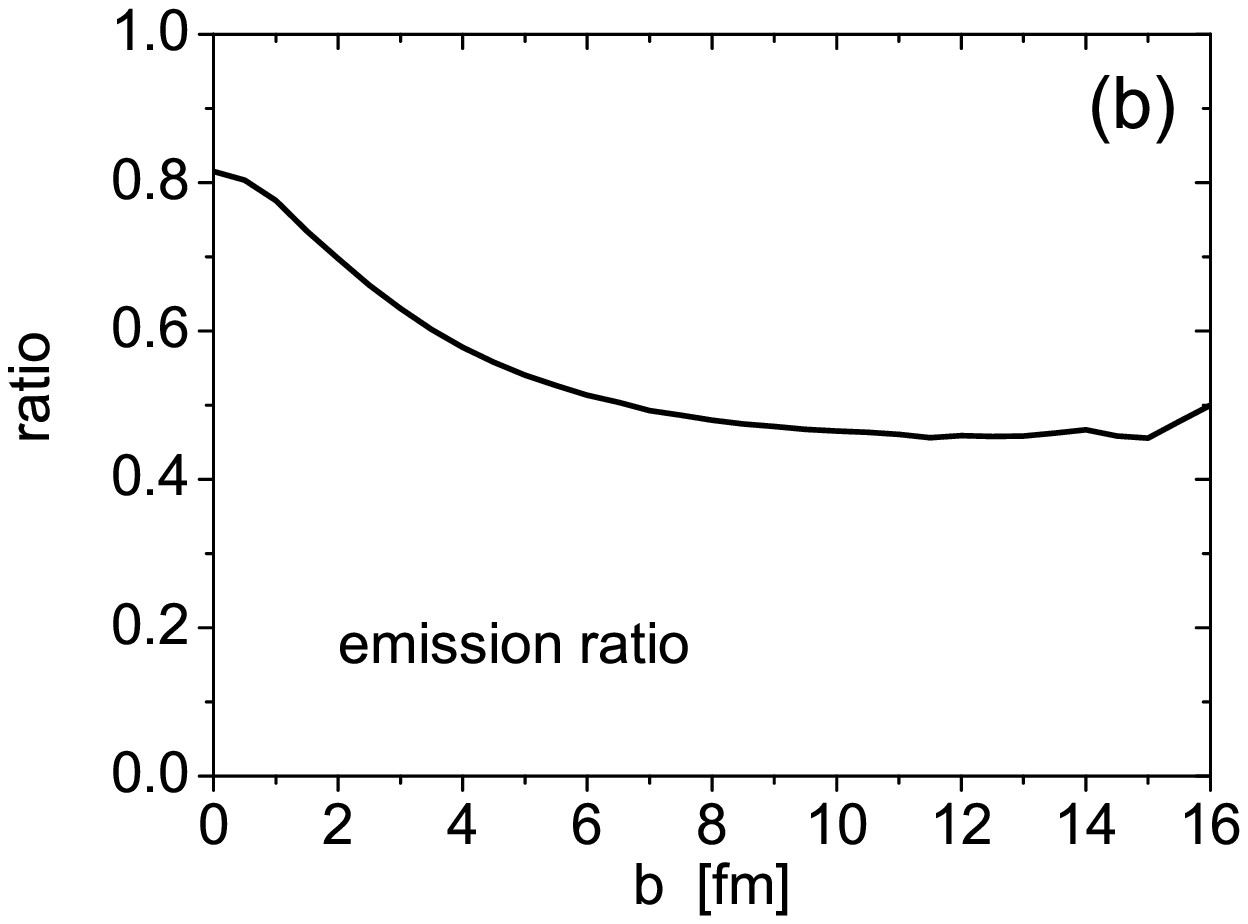}}
\caption{(Color online) (a) The asymmetry defined in Eq. (\ref{asymdef})
  within the approximations described in Sec. \ref{sec:momdis}; the curve labeled
  with \textit{Total} correspond to the inclusion of both correlated and soft nucleons
  from the surface, as well as elastic scattering, absorption and soft nucleons from
  evaporation. Correlations, soft nucleons from the surface and elastic scatterings
  exhibit a strong asymmetry, while soft nucleons from evaporation are emitted
  isotropically.
  (b) The fraction of emitted high-momentum nucleons which survive
  after scattering off the spectator system, disregarding any angular dependence.
}\label{Fig12}
\end{figure*}
\subsection{Evaporation mechanism}\label{sec:evap}

A standard mechanism for description of the spectator emission in the process of fragmentation 
is the excitation of the residual system and subsequent evaporation of nucleons with the
exponential probability for the emitted nucleon kinetic energy $T_n=p^2_n/2m$
\beq
\label{softT}
P(T_n)\,=\,e^{-T_n/T_0}\,,
\eeq
where $T_0$ is typically assumed to be close to the average kinetic energy for the Fermi gas 
\cite{Feshbach:1974af}. We can estimate the kinetic energy $T_0$ for a given value of the impact 
parameter in the following way. We have calculated the energy released by destruction of 
nucleon-nucleon pairs in Sec. \ref{sec:pot}. This energy must be distributed among the
evaporating nucleons according to the following energy balance equation:
\beq
\label{nrgbal}
N_{nucl}\left(\,T_0\,+\,\epsilon\right)\,=\,-V^\prime\,-\,T_{emit}\,,
\eeq
where $N_{nucl}$ is the number of evaporated soft nucleons and $T_{emit}$ is the kinetic energy of
the correlated and uncorrelated nucleons emitted from the surface which we discuss in Secs.
\ref{sec:corr} and \ref{sec:surf}.
The renormalization of $V$ in the right hand side $V^\prime\,=\,V\,(1\,+\,\epsilon/<V>_N)\,<\,0$
reflects the energy released, as described in Sec. \ref{sec:pot}, corrected for the binding 
effects of the removed nucleons.
Equation (\ref{nrgbal}) states that the amount of energy $\epsilon$ must be spent for each
of the emitted nucleons, and the available energy is provided by the calculated potential
energy; note that $N_{nucl}$, the number of emitted nucleons, and $V$ depend on the $AA$ impact
parameter $b$, as does the energy per emitted soft nucleon, $T_0$.

The energy balance expressed by Eq. (\ref{nrgbal}) must be corrected for several additional
effects. First, we have to take into account that a significant fraction of the nucleon 
spectators produced in the decay of the spectator system are bound in light $A=2 \div 4$ and 
heavier nuclear fragments. As a result one has to replace in Eq. (\ref{nrgbal}) 
$N_{nucl}$ by $N_{eff} = N_{nucl}\,+\,K\,N_{frag}$ where $N_{frag}$ is the
total number of nucleons bound in the $A\ge 2$ fragments. To determine this factor we use  
the measurements of Refs. \cite{Appelshauser:1998tt} and \cite{Anderson:1982jx}. 
We extracted from the data what fraction of the nucleons
belongs to the fragments and found it to be  $\approx$ 30\%, leading to $K\approx 0.3$.
Hence we did not include it explicitly in the energy balance of Eq. (\ref{nrgbal}).
(The experiments were performed using light projectiles, and neutron production was not 
measured. We made a natural assumption that for collision of light nuclei the proton and 
neutron spectra coincide. We also assumed that the fraction of energy in the fragments 
remains approximately the same for collisions of heavy nuclei.)

There are other effects to be accounted for in Eq. (\ref{nrgbal}) leading to a modification 
of the initial transferred energy to be eventually available for evaporation. The final 
excitation energy is obtained starting with the initial transferred energy calculated by 
considering the broken potential bonds due the removal of nucleons, as described in Sec. 
\ref{sec:pot}. In  Eq. (\ref{newnrg}) we  kept track of contributions which decrease the 
available energy. Namely we took into account that if a correlated or uncorrelated nucleon 
is emitted from the region close to the interaction surface, its kinetic energy is subtracted 
from the available energy; on the other hand, if the same nucleon is absorbed by the spectator 
system while propagating through it, the corresponding energy is put back into the total. 
We included also the contribution to the evaporation energy of the nucleons produced in the 
primary elastic scatterings which were absorbed by the spectator system. The energy per 
evaporating nucleon $T_0$ is then obtained by replacing Eq. (\ref{nrgbal}) with
\beq
\label{newnrg}
N_{eff}\left(\,T_0\,+\,\epsilon\right)\,=\,-V^\prime\,-\,T_{emit}\,+\,T_{abs},
\eeq
with $T_{emit}$ and $T_{abs}$ the total energies due to nucleon emission and absorption
in the final state, respectively.

The same procedure is used to determine the momentum distributions of produced nucleons.
When nucleons are emitted from the interaction surface, they are assigned a momentum $k$
\begin{figure*}[!htp]
\vskip -0.2cm
\centerline{\includegraphics[height=7.0cm]{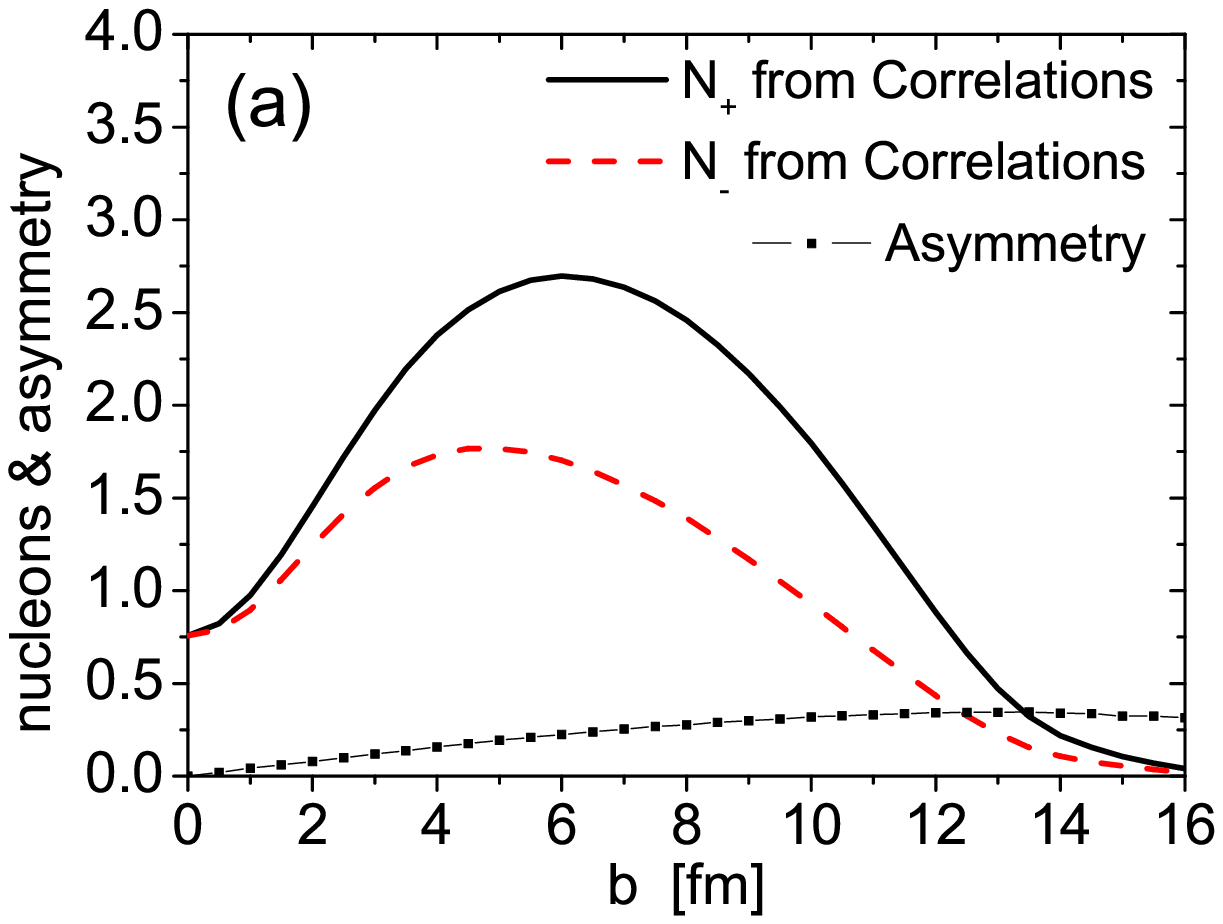}
  \includegraphics[height=7.0cm]{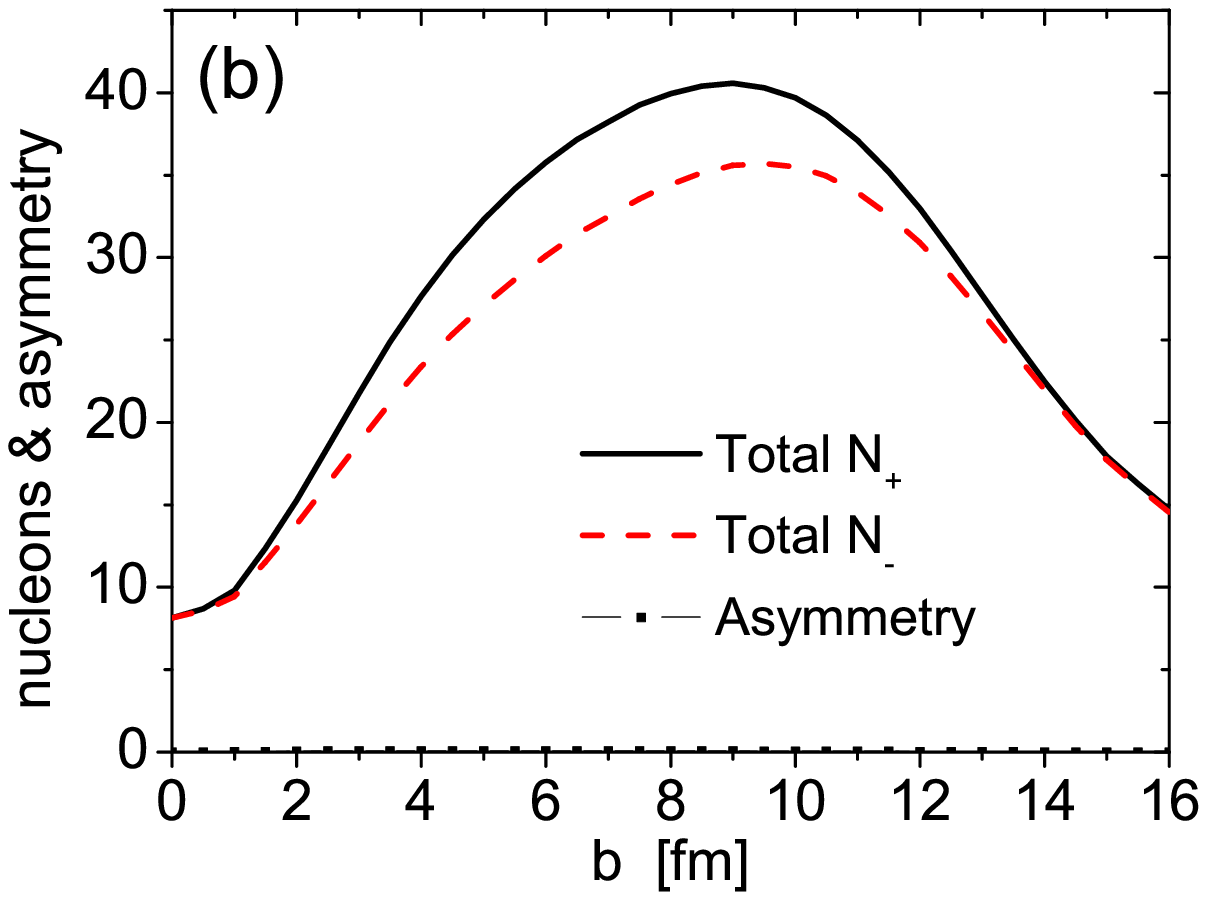}}
\caption{(Color online) The number of high-momentum nucleons emitted in the hemisphere
  containing the projectile ($N_{+}$; solid curve) and in the opposite one
  ($N_{-}$; dashed curve); line and symbol curve represent the
  asymmetry defined in Eq. (\ref{asymdef}).
  (a) Only correlated nucleons;
  (b) quantities calculated using our model with all the different mechanisms
  of nucleon emissions, including soft nucleons from evaporation, as described in
  Sec. \ref{sec:momdis}.
}\label{Fig13}
\end{figure*}
with probability distribution $n_0(k)$ or $n_1(k)$ if they are uncorrelated or correlated, 
respectively. If  they survive propagation through the spectators their momentum is collected. 
The same algorithm is used for the nucleons experiencing the elastic scattering in the initial 
state. These nucleons  are assigned initial momenta  with probability $n_0(k)+n_1(k)$.
Momentum distribution and other results will be presented in the following.
\subsection{Results of calculations}\label{sec:res}

Results for the calculation of the momentum distributions are shown in Figs. \ref{Fig7} 
and \ref{Fig8}. The curves correspond to the inclusion of the different effects we have 
described in this section. The curve labeled Evaporation corresponds to the energy per 
emitted nucleon obtained by dividing the total potential energy by the number of 
the expected free nucleons. The curve labeled with Total is obtained from Eq. 
(\ref{newnrg}), where we account for (i) the energy going into the kinetic energy of the 
emitted particles $T_{emit}$ and the kinetic energy of the absorbed particles $T_{abs}$ going 
into additional available energy and (ii) the number of nucleons emitted in the first stage 
from the surface close the interaction. These are either high-momentum or soft nucleons. 
The number of nucleons emitted from the surface is subtracted from the number of free 
nucleons which are produced in the evaporation. The curves labeled Correlation
correspond to the inclusion of high-momentum-correlated nucleons; curves Soft
correspond to the inclusion of soft nucleons emitted from the surface, which are the 
nucleons left after subtracting the correlated from the active ones; the curves labeled 
Elastic correspond to the contributions from primary elastically scattered nucleons.
Note that subsequent absorption from the residual system by elastic rescattering is taken 
into account for all the separate contributions. The results of our calculation of the energy 
per evaporating nucleon are  shown in Fig.~\ref{Fig9} as a function of the impact parameter 
[\ref{Fig9}(b)]. Both the result including only   evaporation and total result [Eq. (\ref{newnrg})] 
are given. Figure \ref{Fig9}(a) shows the dependence of the number of interacting and spectator 
nucleons on $b$ of the number of spectators.

We would like to comment on data from Ref. \cite{Anderson:1982jx}. Momentum
distributions of detected nucleons were explicitly measured; nonetheless, a direct
comparison with our results may not be trivial. The authors have compared the observed
momentum distribution with the theoretical momentum distribution in nuclei, which is not
consistent with the observation that (i) most of the nucleons are emitted by the evaporation
mechanism, whose distribution is given by Eq. (\ref{softT}) and it is expected to differ
significantly from the momentum distribution of nucleons inside the nucleus and (ii) the
observed momentum distribution depends on the transferred energy in the process of removing 
one nucleon while the momentum distribution is the integral of the spectral function for one
nucleon removal over all the removal energy range; a calculation of the effects of the
integration of the spectral function on a limited region of energy may be found in Ref.
\cite{ciofipacesalme}. (iii) We suggest that only the high-momentum part of the observed
momentum distribution can be compared with the nucleon momentum distribution in nuclei 
$n(k)$, since it is due to direct emission of correlated nucleons whose momentum may be 
distorted by the
propagation through the spectator system but still will be similar to the original $n(k)$. 
Moreover, the high-momentum region of the observed momentum distribution is also affected 
by primary elastically scattered nucleons, which must be taken into account. Data from Ref.
\cite{Anderson:1982jx} exhibits a change in slope of the momentum distribution of protons 
as well as a strong change of the shape of the transverse-momentum distribution for nucleons 
with total energy equal to $E_A/A$. The first effect  may be due to the high-momentum, 
correlated nucleons, as suggested by results of calculations shown in Fig.~\ref{Fig7}
\textit{(b)}. While the second effect is likely to be due to the elastic-scattering mechanism.
\subsection{Sensitivity to the parameter $r_{max}$}\label{sec:mdep}

Here we return to discussion of $r_{max}$ and sensitivity of the results to a specific
choice of $r_{max}$ which played an important role in identifying the \textit{active} nucleons.
We already gave our rationale for the choice of $r_{max}\approx 2 fm$. An alternative way to
determine $r_{max}$ is to require that it is equal to the average maximum distance from the
nearest interacting nucleons of spectators with potential energy larger than the average
$<V>/A$. This definition leads to the value of  $r_{max}=2 fm$, which is close to our qualitative
expectations. To explore the sensitivity to this parameter we varied it between 1.8 and 2.2
fm. The results are shown in Fig.~\ref{Fig11}. It can be seen that the final result for the
energy per evaporating nucleon [\ref{Fig11}(a)] changes by at most 10\%, with an even smaller
effect on the momentum distribution for $k < 300$ MeV/$c$ [\ref{Fig11}(b)]. The effect is larger for
high momentum nucleons. However, it should be stressed that reducing $r_{max}$  to less than 2 
fm would produce an unrealistically low number of correlations, since the $NN$ correlation 
functions of Eq. (\ref{corfs}) have been shown by many-body calculations (see the correlation 
functions in Refs. \cite{Alvioli:2005cz} and \cite{Geurts:1996zz} and references therein) to extend
up to relative distances of at least 2 fm. Also, we checked that the variation of $r_{max}$ leads
to a very small modification of the asymmetry of emission, which we discuss in the next section.
\section{Angular dependence of directed flow of emitted high-momentum nucleons}\label{sec:angular}

One of the aims of this work is to estimate the asymmetry of emerging from the $AA$ collision.
These nucleons are generated on the inner surface left after the fast propagation of the 
projectile nucleus; we define $N_{+}$ to be the number of nucleons ending up in the hemisphere 
oriented toward free space and $N_{-}$ its analog in the opposite hemisphere.
We can define the impact parameter-dependent asymmetry as follows:
\beq
\label{asymdef}
A(b)\,=\,\frac{N_{+}\,-\,N_{-}}{N_{+}\,+\,N_{-}}\,.
\eeq
We expect that the few nucleons
surviving in a central collision will find little or no matter to propagate through
in any direction, resulting in $A(0)=0$. The situation is depicted in Fig.~\ref{Fig10},
for different impact parameters ranging from $b=1$ fm (top left panel) to $b=15$ fm
(bottom right panel). In the first case of almost central collisions, few nucleons are
left; they are far from each other and free to propagate. For increasing $b$, it can be seen
how an asymmetry arises in the rest frame of the target nucleus, since nucleons are
produced with random momentum direction and they can end up propagating into free
space or through the spectator system which prevents emitted nucleons to propagate
freely.
For a large impact parameter, the asymmetry of Eq. (\ref{asymdef}) which is due only to
the  correlations should approach the value given by taking $N_{-}\simeq 0.5N_{+}$,
which is the fraction of nucleons surviving after the scatterings through the spectator
matter at large impact parameters. This would result in $A(b)\simeq 0.3$. The calculations
of $A(b)$ performed within our model are shown in Figs. \ref{Fig12} and \ref{Fig13} and confirm
these expectations. Figure~\ref{Fig12} shows the asymmetry including nucleons from the different
emission mechanisms. The contribution due to correlations is, as already stated, strongly 
asymmetric; elastically scattered nucleons lead to similar asymmetry as they also originate 
from the surface; soft nucleons originating from the surface somewhat increase asymmetry. 
The largest contribution, as far as the number of nucleons is concerned, is from the soft, 
evaporating ones, and makes the total asymmetry much smaller. Figure~\ref{Fig13} shows the 
individual contributions to $N_+$ and $N_-$ from correlated nucleons alone and from the 
total number of emission of spectator nucleons. Note that for $b > 12$ fm we used an 
extrapolation of the NA49 data \cite{Appelshauser:1998tt} for the number of nucleons 
produced at a given impact parameter for determining the fraction of energy taken from 
nucleons and fragments, as described in Sec. \ref{sec:momdis}.

In this analysis we did not take into account the effect of the Coulomb interaction between nuclei
which gives fragmenting systems of opposite transverse momenta. The effect gives an asymmetry of
the opposite sign than the effect we discussed here. Our effect is large for large nucleon momenta,
while the Coulomb effect is most important for the lowest momentum neutrons. We will consider the
Coulomb effect elsewhere.

We mentioned in the Introduction that RHIC experiments observed displacement of the core of the
neutron shower from the center of the detector. However, this effect is dominated by low-momentum
neutrons and may be more sensitive to the Coulomb effect. Hence it would be important to study
asymmetry separately for different ranges of the neutron momenta as well as study experimentally
correlation of the strengths and directions of asymmetries of two fragmentation regions.
When the origin of the asymmetry is cleared out it would be possible to use this effect to solve
the ambiguity in the sign of the impact parameter $\Vec{b}$. Note here, that the  correlation of
$v_1$ with corresponding asymmetry for hadrons at the rapidities away from the central one was
observed at RHIC. However due to the lack of understanding of the origin of the neutron asymmetry
it was impossible to use this observation to constrain the current models of $AA$ collisions at RHIC.

We investigated the dependence of the $A(b)$ on the cutoff distance from the closest interacting
nucleon, as discussed in the previous section. We find negligible effects on the asymmetry. This
is due to the fact that the emission of correlated nucleons is highly asymmetric, but the asymmetry
is dominated by isotropic emissions in any case so the small relative fraction of correlations
practically does not affect $A(b)$.

\section{Comparison with the abrasion-ablation model}\label{sec:abl}

In this section we compare our novel approach for the calculation of the excitation energy
of the spectator system with the one employed in Ref. \cite{Scheidenberger:2004xq}. In the
cited Ref. \cite{Scheidenberger:2004xq}, the authors use the abrasion-ablation model for 
hadronic interactions and
the relativistic electromagnetic dissociation model for electromagnetic interactions of
relativistic heavy ions to describe data of charge-changing cross section in Pb-$A$ collisions.
In particular they address the question of how to estimate the excitation energy of a nuclear
system formed by sudden removal of several nucleons, which is also one of the main aims of the
present work, and describe the decay of excited nuclear systems within the statistical
multi-fragmentation model. The abrasion model describes participant and spectator nucleons
with participants originating from the overlapping parts of the colliding nuclei, while their
non-overlapping parts are treated as spectators which represent excited remnants of the initial
nuclei which undergo secondary decay by statistical evaporation and fission models in the
so-called ablation model.

The $NN$ interaction probability in Ref. \cite{Scheidenberger:2004xq} is defined through
(uncorrelated) nuclear thickness functions, while in our approach correlations are taken
into account automatically by using improved configurations from Ref. \cite{Alvioli:2009ab}.
The main difference then consists in the estimate of the spectator system (prefragment, in
the terminology of Ref. \cite{Scheidenberger:2004xq}) excitation energy. We use similar 
values of $NN$ cross sections. The authors of Ref. \cite{Scheidenberger:2004xq} describe 
excitation energy by the abrasion model, which basically evaluate the energy due to a 
hole in the uncorrelated ground state of the initial nucleus, while we use the realistic 
calculation of Sec. \ref{sec:pot}; moreover, in Ref. \cite{Scheidenberger:2004xq} it is 
explicitly mentioned that the procedure is not well defined for a large number of removed 
nucleons and that their method of calculating excitation energies of prefragments 
via the hole state densities should be considered only as a model assumption.
A comparison with other approaches was also done in Ref. \cite{Scheidenberger:2004xq},
finding for the excitation energy per nucleon a value of $\simeq 40$ MeV using the ablation
model and $\simeq 27$ MeV using a more refined formula, the latter estimate being confirmed
by data; we note that the value of $\simeq 27$ MeV is in good agreement with our findings of
Fig.~\ref{Fig9}, where this value is just about our estimate for $b\simeq 3$-4 fm; we believe
that one of the important improvement of our approach is precisely the possibility of giving
an impact-parameter dependent estimate of the excitation energy, which was untouched by any
of the previous approaches.
As a last remark, let us stress that in Ref. \cite{Scheidenberger:2004xq} it was mentioned that
their model should be complemented with FSI, which we take into account in our estimate of the
energy per nucleon, and the additional energy brought in by elastically scattered primary nucleons
subsequently absorbed by the spectator system, which we also considered.

\section{Conclusions}\label{sec:conc}
We have demonstrated that the underlying dynamics of the process of the nucleus fragmentation in
high energy nucleus-nucleus collisions is strongly related to the  the presence of short-range
correlations in nuclei. We predict a number of new phenomena which could be tested in the current
and forthcoming heavy ion experiments, including the dependence of the momentum spectrum on the impact
parameter strong asymmetry of the emission nucleons along the $b$ direction. Such an asymmetry as well
as other predicted effects may be of use for more detailed analyses of dynamics of  the heavy ion 
collisions. In particular, it would allow one to investigate whether similar asymmetry is present for  
the hadrons produced in $AA$ collisions away from zero center of mass rapidity.
We also predict a close connection of the spectrum of nucleons in the central collisions and momentum
distribution in the nuclei. We are now in the process of implementing the discussed effects in a complete
MC event generator of $AA$ collisions and results will be presented elsewhere \cite{alvnew}.

\section{Acknowledgements}\label{sec:ack}

We thank G.~Bertsch, T.~Csorgo, L.~Frankfurt, I.~N.~Mishustin, I.~A.~Pshenichnov, M.~Vargyas,
S.~Voloshin and S.~White, for very useful discussions. This work is supported by a DOE grant under
contract number DE-FG02-93ER40771. During substantial revision of the manuscript M.A. was supported by
the project HadronPhysics2 of the FP7 of the EU (Grant No. 227431). M.A. thanks the HPC-Europa2
Consortium (project number: 228398), with the support of the EC - Research Infrastructure Action 
of the FP7 - and EPCC, Edinburgh for the use of computing facilities.


    \end{document}